\documentclass{emulateapj}
\usepackage{apjfonts}
\usepackage{amsmath}

\newcounter{ichi}
\newcounter{ni}
\newcounter{san}
\newcounter{yon}
\makeatletter
\newsavebox{\@parc@ption}
\def\parcaption#1{%
\sbox{\@parc@ption}{\shortstack[l]{#1}}%
>\setbox\@tempboxa\hbox{\csname fnum@\@captype\endcsname}%
\@tempdima\columnwidth \advance\@tempdima-\wd\@tempboxa
\@tempdimb\@tempdima 
\ifdim\wd\@parc@ption>\@tempdimb \@tempdima\@tempdimb
\else\@tempdima\wd\@parc@ption\fi
\sbox{\@tempboxa}{\parbox[t]{\@tempdima}{#1}}%
\caption{\usebox{\@tempboxa}}}
\makeatother

\slugcomment{}

\shorttitle{
High-Energy GRB Afterglows in the Early and Late Jet Model 
}
\shortauthors{Murase et al.}

\begin{document}
\title{
On the implications of late internal dissipation for shallow-decay afterglow emission and associated high-energy gamma-ray signals
}


\author{Kohta Murase\altaffilmark{1,2}, Kenji Toma\altaffilmark{3},  
Ryo Yamazaki\altaffilmark{5} and Peter M\'esz\'aros\altaffilmark{3,4}}


\altaffiltext{1}{Center for Cosmology and AstroParticle Physics, Ohio State University, 191 West Woodruff Avenue, Columbus, OH 43210, USA}
\altaffiltext{2}{Department of Physics, Tokyo Institute of Technology, 2-12-1 Ookayama, Meguro-ku, Tokyo 152-8550, Japan}
\altaffiltext{3}{Department of Astronomy \& Astrophysics, Pennsylvania State University, University Park, PA 16802, USA}
\altaffiltext{4}{Center for Particle Astrophysics, Pennsylvania State University, University Park, PA 16802, USA}
\altaffiltext{5}{Department of Physics and Mathematics, Aoyama Gakuin University, 5-10-1 Fuchinobe, Sagamihara 252-5258, Japan}

\begin{abstract}
The origin of the shallow-decay emission during early X-ray afterglows has been an open issue since the launch of the \textit{Swift} satellite. One of the appealing models is the late internal dissipation model, where X-ray emission during the shallow-decay phase is attributed to internal dissipation, analogous to the prompt gamma-ray emission.  We discuss possible scenarios of the late prompt emission, such as late internal 
shocks, magnetic reconnection, and photospheric dissipation. We also consider the consequences of late dissipation and a two-component (early and late) jet model for the high-energy (GeV-TeV) emission. We study not only synchrotron self-Compton (SSC) emission from the early and late jets but also external inverse-Compton (EIC) emission, which is naturally predicted in the late dissipation model. For the latter, we perform numerical calculations taking into account the equal-arrival-time surface of EIC photons and show that the EIC component typically has a peak at $\sim 1-100$~GeV which may dominate over the SSC components. 
We demonstrate that very high energy gamma rays from both these components are detectable for nearby and/or energetic gamma-ray bursts, with current and future Cherenkov detectors such as MAGIC, VERITAS, CTA and HAWC, and possibly \textit{Fermi}. Although the expected event rate would not be large, detections should be useful as a test of the model. Multi-wavelength observations using both the ground-based telescopes and the \textit{Swift} and/or \textit{Fermi} satellites are also important to constrain the models.
\end{abstract}

\keywords{gamma-rays burst: general --- radiation mechanisms: non-thermal}

\section{Introduction} 
An understanding of the mechanism controlling the early X-ray afterglow emission from gamma-ray bursts (GRBs) has been one of the most debated issues since the launch of the \textit{Swift} satellite. 
The canonical X-ray afterglow can be classified into three phases: the steep-decay phase~\citep[see, e.g.,][]{KP00,YTIN06,Zha+06}, the shallow-decay (or plateau) phase and the normal-decay phase~\citep[see, e.g.,][and references therein]{Nou+06,OBr+06,Pan+06,Zha+06,Wil+07}. 
In particular, the shallow-decay phase is difficult to explain by the standard external forward shock (FS) model~\citep[see reviews, e.g.,][]{Mes06,Zha07}. Numerous models have been proposed so far to explain it. 
Most frequently discussed models are modified external shock models. One of the most popular interpretations involves either a continuous energy injection into the external FS, where the long-lasting central engine energy output has a smooth decline $\propto T^{-q}$, or short-lived central engine ejects shells with a steep power-law distribution of bulk Lorentz factors which can explain the shallow-decay emission~\citep[e.g.,][]{RM98,DL98,ZM01a,Zha+06}. Another version invokes a time-dependent microphysical 
scenario in the FS model, where changing parameters such as $\epsilon_e$ lead to the observed shallow-decay emission~\citep[e.g.,][]{ITYN06}. Dermer (2007) showed that the shallow decline may be explained if ultrahigh-energy cosmic rays are efficiently produced, by the recovery of an adiabatic relativistic blast wave after its radiative phase due to efficient photomeson losses and particle escape. The external reverse shock (RS) may also account for the shallow-decay emission, with an appropriate $\Gamma$ distribution of the ejecta, if the RS emission dominates over the FS emission in the X-ray band~\citep[e.g.,][]{GDM07,UB07}. 
On the other hand, some authors have suggested explanations based on two- or multi-component jet scenarios. 
Panaitescu (2008a) showed that upscattering of FS photons by a relativistic shell can 
outshine the standard FS emission. Recently, Yamazaki (2009) proposed an alternative interpretation, where X-ray light curves are explained by the difference between the X-ray onset time and the burst trigger time. Among geometrical models, two co-aligned jets with different opening angles, i.e., wide and narrow jets, can also lead to the shallow-decay emission~\citep[e.g.,][]{EG06}, while a multiple-sub-jets model was also 
proposed as one of the explanations~\cite{TIYN06}.  
 
 Another attractive interpretation is that X-ray emission is attributed to long-lasting internal 
dissipation~\cite{Ghi+07,Kum+08b}. This long-lasting dissipation or late prompt emission model can explain the chromatic behavior, which is not so easy to explain in modified external shock models where the optical flux would presumably track trends of the X-ray flux. This model is viable in the sense that the shallow-decay phase is not ubiquitous and some GRB afterglows are explained simply by the standard FS model when the late prompt emission is weak. Especially, some of the GRBs such as GRB 070110 have a plateau and a following steep decline, which strongly suggests that X-rays originate from the late internal dissipation rather than the FS emission~\citep[e.g.,][]{Tro+07,LZZ07}. This late internal dissipation model may also be consistent with the existence of X-ray flares in the early afterglow phase~\citep[e.g.,][]{Fal+07,Chi+10}.  However, the situation is still inconclusive and unclear. For example, the lack of spectral evolution across the transition from the plateau to the normal-decay phase and the compliance of the closure relations in the normal-decay phase after the transition may rather suggest FS models~\cite{LZZ07,Lia+09}. 
  
On the other hand, recent novel results from the \textit{Fermi} satellite have provided us with 
interesting clues for the mechanisms of GRBs. Especially, the onboard Large Area Telescope (LAT) has detected high-energy ($>$ GeV) gamma rays from a fraction of GRBs~\cite{Abd+09a,Abd+09b,Ack+10}. Those detections have provided not only clues to the prompt emission mechanism but have also provided the first detailed data about the high-energy afterglow emission, which had been expected for many years~\citep[e.g.,][]{DCM00,SE01,ZM01a}. In fact, late-time high-energy gamma-ray emission from GRBs such as 080916C, 090510, and 090902B has been attributed to afterglow emission rather than the prompt emission~\cite{KD10,Ghi+10,He+10}. 
Various theoretical possibilities for the high-energy emission mechanism have also been discussed by numerous authors. The most widely discussed mechanisms are synchrotron and synchrotron self-Compton (SSC) emission~\citep[see review, e.g.,][and references therein]{FP08}. The external inverse-Compton (EIC) emission 
has been considered in some cases where seed photons come from flares or prompt 
emission~\citep[e.g.,][]{Bel05,Wan+06,Pan08b,Mur+10}. If protons and nuclei are accelerated up to very high energies, hadronic gamma-ray afterglows are also expected via the photomeson production and ion synchrotron radiation~\citep[e.g.,][]{BD98,PW05,Mur07}.  

Despite recent progress in observations of high-energy gamma rays, the link between GRBs with shallow-decay emission and GRBs whose high-energy emission is detected by \textit{Fermi} is uncertain, since due to the scarcity of simultaneous detections with \textit{Swift} it is unclear whether GRBs detected by 
\textit{Fermi}/LAT do have the shallow-decay phase or not. Also, the detectability with \textit{Fermi} is limited at late times so that it is not easy to distinguish among the various models for shallow-decay emission. In this sense, Cherenkov detectors such as MAGIC and VERITAS may be more important. Although detections of $> 10$~GeV photons from distant GRBs become difficult because of the attenuation by the extragalactic background light (EBL), Cherenkov telescopes could provide many more photons than \textit{Fermi} when a nearby and/or energetic burst occurs. Although very high-energy photons from GRBs have not been firmly detected so far, the future CTA and AGIS arrays would significantly increase the chances to observe high-energy GRB emission.   

Given a high enough detection rate of high-energy photons by such observatories, high-energy gamma rays would provide very useful probes of  the origin of shallow-decay emission. For example, Murase et al. (2010) demonstrated that EIC emission would be important to diagnose the prior emission model~\cite{Yam09}, which 
is one of the two-component (early and late) jet models.  In this work, motivated by the above prospects, we discuss theoretical possibilities of late internal dissipation scenarios and investigate the associated high-energy emission. First, we review various late prompt emission scenarios, such as the late internal shocks, dissipative photosphere, and magnetic dissipation scenarios. Second, we analytically study the high-energy gamma-ray emission expected in the internal dissipation model. In particular, we numerically calculate the EIC emission in detail, which plays an important role in two-component jet models such as the late prompt emission model. Throughout this work, cosmological parameters are set to $H_0=71~{\rm km}~{\rm s}^{-1}~{\rm Mpc}^{-1}$, $\Omega_M=0.3$, and $\Omega_\Lambda=0.7$, and we adopt the conventional notation $Q=Q_x \times {10}^{x}$.  

\section[]{Theoretical Possibilities of Late Prompt Emission}
Observationally, a good fraction of GRB afterglows show shallow-decay emission from $T \sim {10}^{2.5}$~s to $T \sim {10}^{3.5}$~s~\citep[e.g.,][]{OBr+06,Wil+07,Lia+09}, which can be expressed as
\begin{equation}
L_{\rm LP}(T) \propto 
\left\{ \begin{array}{ll} 
T^{-\alpha_{\rm fl}}
& \mbox{($T < T_{a}$)}\\
T^{-\alpha_{\rm st}}
& \mbox{($T_a \leq T$)}
\end{array} \right.
\end{equation} 
Here $T_a \sim {10}^{3}$~s is the break time when the shallow-decay phase ceases, $\alpha_{\rm fl} \sim 0.0-0.5$ and $\alpha_{\rm st} \sim 1.0-2.0$. Many GRBs show the chromatic behavior, where optical and X-ray afterglows evolve in different ways, which tempts one to consider a two-component interpretation, i.e., X-ray and optical emissions come from different emission regions. Some authors argued that the shallow-decay X-ray 
emission may be attributed to emission caused by internal dissipation similar to that of the prompt 
emission, while the normal-decay optical emission is interpreted as an external FS component~\cite{Ghi+07,Kum+08b}. 
For example, Ghisellini et al. (2009) successfully fitted X-ray and optical afterglows of various bursts in this picture. The isotropic radiation energy of late prompt emission, $\mathcal{E}_{\rm LP}^{\rm iso}$, is typically $\mathcal{E}_{\rm LP}^{\rm iso} \sim (0.01-0.1) \times \mathcal{E}_{\rm GRB}^{\rm iso} \sim {10}^{51-52}$~erg, where $\mathcal{E}_{\rm GRB}^{\rm iso}$ is the isotropic radiation energy of prompt emission. 

Such an interpretation seems strongly supported for a fraction of bursts. Some GRB afterglows show even a plateau rather than a shallow decay. For example, GRB 070110 has the plateau, $\alpha_{\rm fl} \sim 0.09$, and the following sudden decline, $\alpha_{\rm st} \sim 9$~\cite{Tro+07}. Such a behavior is very difficult to explain in the context of modified external shock models, which rather reflect variable activities of the long-lived central engine, although some of the issues can be solved if the emission is anisotropic in the comoving frame~\cite{Bel+11}. Then, the late prompt emission would presumably be attributed to long-lasting internal dissipation. 
This interpretation seems consistent with the existence of flares, which are observed in about $\sim 30-50$~\% of GRBs~\cite{Fal+07,Chi+10}, and flares may be attributed to accidental events of stronger internal dissipation, in the late internal dissipation scenario~\cite{Ghi+07,Kum+08b}. The duration of flares $\Delta T_{\rm flare}$ is shorter than the observation time $t$ and its flux enhancement is striking (the energy fluence is about 10~\% of prompt emission), which suggests that they originate from temporarily strong late internal dissipation by the long-lasting central engine~\cite{IKZ05}. One puzzling point is that the pulse width of flares increases linearly with time~\cite{Chi+10}, while those of prompt emission have various durations, showing no increasing pulse width during the burst. Possibly, this may reflect the behavior of the central engine. 
In the late internal dissipation model, one may expect some time variability in the apparently smooth X-ray light curve of the shallow-decay emission, where it may also include some information on the central engine. However, at present, it is difficult to measure well even if the emission consists of numerous events of small internal dissipation~\citep[e.g.,][]{Ghi+07}.

If the shallow-decay emission originates from long-lasting internal dissipation, what activity of the central engine could be responsible for it? For the central engine of GRBs, two possibilities have been most frequently discussed, accretion of matter onto a black hole~\cite{Kum+08a} or a fast rotating magnetar~\citep[e.g.,][]{DL98,ZM01a,TCQ04,YCC10}. In the latter scenario, the break time $T_{a}$ can be attributed 
to the spin down time. 
\begin{equation}
L_{\rm LP} (T) \propto L_{P} (T)
\propto
\left\{ \begin{array}{ll} 
{\rm const.}
& \mbox{($T < T_a$)}\\
T^{-2}
& \mbox{($T_a \leq T$)}\\
\end{array} \right.
\end{equation}
where $L_P$ is the spin down luminosity. In this scenario, the outflow would initially be Poynting-dominated. 

In the former scenario, the shallow-decay behavior is attributed to the activity of the system of a black hole with an accretion disk, e.g., mass fall back accretion onto the central black hole. The break time $t_a$ can be interpreted as the end time of mass fall back. Kumar et al. (2008b) proposed that prompt emission is associated with the accretion of the innermost region of the progenitor star, whose angular velocity is small, while the outer envelope with the larger angular velocity is responsible for the shallow-decay emission. The outflow luminosity is expected to be proportional to the mass accretion rate, and then the temporal index $\alpha_{\rm fl}$ is related to its behavior. After $T_{a}$, we expect~\cite{Kum+08a}
\begin{equation}
L_{\rm LP}(T) \propto \dot{M}_{\rm BH} (T) \propto 
{\left[1+\frac{3}{2s-1}\frac{T-T_a}{T_{\rm ac}} \right]}^{-\frac{4(s+1)}{3}} 
\end{equation}
where $T_{\rm ac}$ is the accretion timescale and $s \sim 0-1$. The above expression can explain both the rapid decline (when $T_{\rm ac}<T_a$) and smooth transition (when $T_a < T_{\rm ac}$) at $T \sim T_a$. 

There is also another interpretation of the behavior after $T_a$ and the origin of $T_a$. If the late jet continuously decelerates, one expects a jet break in observed light curves when $\Gamma_{\rm LP}$ becomes $\theta_{\rm LP}^{-1}$~\cite{Ghi+07}. 

Note that the late internal dissipation could also explain the steep-decay emission just after the prompt emission phase, although it is usually attributed to the high-latitude prompt emission~\citep[e.g.,][]{KP00,YTIN06}. For example, in the collapsar scenario, Kumar et al. (2008b) suggested that the accretion of gas from the ``transition'' region between the core and the envelope, where the density has a steep decline, may lead to the steep-decay emission. However, such possibility that X-ray tails reflect the dying history of the central engine totally depends on models of the central engine, and the present situation is unclear since the apparent spectral evolution is also affected by the intrinsic spectrum of the prompt emission~\cite{Zha+09}.
  
At present, it is difficult to discriminate among these possibilities from observations. Therefore, for 
the discussions below, we treat the temporal indices $\alpha_{\rm fl}$ and $\alpha_{\rm st}$ as 
just parameters determined from observations. Also, we assume that the long-lasting internal dissipation occurs according to Equation~(1) without specifying the central engine.     

\subsection{Late Internal Shock Scenario}
In the classical scenario, the prompt emission is explained by electromagnetic radiation from electrons accelerated at internal shocks that occur in the optically thin relativistic outflow~\cite{RM94}. 
Flares may also be explained similarly, where X-ray and/or ultraviolet photons are produced by relativistic electrons accelerated at late internal shocks~\cite{FW06}. The bulk Lorentz factor responsible for flares is often thought to be smaller than that of prompt emission~\citep[e.g.,][]{JFW10}, and then, applying this scenario to late prompt emission, the typical collision radius is estimated as $r_i \approx 2 \Gamma_{\rm LP}^2 {\delta T}_{\rm var} \simeq 1.5 \times {10}^{15}~{\rm cm}~{(\Gamma_{\rm LP}/5)}^2 {\delta T}_{\rm var,3} {(1+z)}^{-1}$. (However, in some models such as the fast rotating magnetar model~\cite{TCQ04}, the bulk Lorentz factor of the late jet may be much larger.)

Let us consider the two-shell collision between fast and slow shells with $\Gamma_{f}$ and $\Gamma_s$, respectively. The relative Lorentz factor between the shells is $\Gamma_{\rm LP,sh} \approx (\Gamma_f/\Gamma_s+\Gamma_s/\Gamma_f)/2 \sim 5$ and the average Lorentz factor of random internal motions is $\Gamma_{\rm LP,is} \approx  (\sqrt{\Gamma_f/\Gamma_s}+\sqrt{\Gamma_s/\Gamma_f})/2 \sim 0.5$ for $\Gamma_f \sim 13$ and $\Gamma_s \sim 2$ (where the Lorentz factor of the merged shell is $\Gamma_{\rm LP} \approx \sqrt{\Gamma_f \Gamma_s} \sim 5$). 
Then, the electron injection Lorentz factor is obtained as
\begin{equation}
\gamma_{e,m} \approx \frac{\epsilon_e}{f_e} (\Gamma_{\rm LP,\rm is}-1) \frac{m_p}{m_e}
\simeq 0.92 \times {10}^{3} \left( \frac{\Gamma_{\rm LP,is}-1}{0.5} \right) \epsilon_{e,-1} f_{e,-1}^{-1},
\end{equation}
where $\epsilon_{e}$ is the fraction of the internal energy transferred to non-thermal electrons and $f_e$ is a number fraction of accelerated electrons. Introducing $\epsilon_{B}$ which is the fraction of the internal energy transferred to the magnetic field, the comoving magnetic field is estimated as $B \simeq 3.5 \times {10}^{3}~{\rm G}~{\left[ \frac{(\Gamma_{\rm LP,is}+3/4)(\Gamma_{\rm LP,is}-1)}{(9/8)}\right]}^{1/2} {\epsilon}_{B,-1}^{1/2} L_{k,49}^{1/2} r_{i,15}^{-1} {(\Gamma_{\rm LP}/5)}^{-1}$, and then the observed synchrotron peak energy is
\begin{eqnarray}
E^b &\simeq& 0.17~{\rm keV}~\left[ \frac{{(\Gamma_{\rm LP,is}-1)}^{5/2} {(\Gamma_{\rm LP,is}+3/4)}^{1/2}}{(3/8\sqrt{2})} \right] \nonumber \\ &\times& f_{e,-1}^{-2} \epsilon_{e,-1}^2 \epsilon_{B,-1}^{1/2} L_{k,49}^{1/2} r_{i,15}^{-1} {(1+z)}^{-1}.  
\end{eqnarray}
The observed X-ray emission shows a hard spectrum in the X-ray band, $F_{\rm LP} \propto E^{-1}$, which is attributed to synchrotron emission by relativistic electrons with the spectral index of $p \sim 2$. 
Accelerated electrons are expected to be in the fast cooling regime.

In the case of prompt emission, the classical scenario has several problems in explaining observations. One of them is that, if the radiation mechanism is synchrotron emission, explaining the low-energy spectral index of $\beta_l \sim 1$ is not easy~\citep[but see, e.g.,][]{DKK01,NAS09,DBD11}. But, since the spectrum in the far-ultraviolet range is not observed, it is uncertain whether the late internal shock scenario suffers from the same issue.
Another one is the efficiency problem that energy transferred to prompt emission often seems larger than the afterglow kinetic energy~\citep[e.g.,][]{Zha+07}. 
The efficiency seems worse due to weaker collisions in the late internal shock model, but it is observationally unclear whether the shallow-decay emission has the same problem.

\subsection{Dissipative Photosphere Scenario}
In the previous subsection, we discussed the late internal shock model based on the analogy to the prompt emission. However, the prompt emission mechanism itself is still under debate, and many possibilities have been suggested. Another popular scenario of the prompt emission is the photospheric emission model, where quasi-thermal emission comes from around the photosphere ($\tau_T=n_e \sigma_T (r_i/\Gamma)\sim1$)~\citep[e.g.,][]{Tho94,MR00,Mes+02,RM05,PMR06}. Although there are various versions~\citep[e.g.,][]{Iok+07,Bel10,Iok10}, 
we here consider the dissipative photosphere scenario~\cite{RM05,PMR06}, where internal dissipation (via e.g., internal shocks or magnetic reconnection) occurs around the photosphere. For the late jet making late prompt emission, the photospheric radius is written as 
\begin{eqnarray}
r_{\rm ph} &=& \left( \frac{\zeta_e L_{\rm LP} \sigma_T}{\epsilon_r 4 \pi \Gamma_{\rm LP}^3 m_p c^3} \right) \nonumber \\
&\simeq& 9.4 \times {10}^{13}~{\rm cm}~(\zeta_e/10 \epsilon_r) L_{\rm LP,48} {(\Gamma_{\rm LP}/5)}^{-3}, 
\end{eqnarray}
where $\epsilon_r$ is the ratio of the radiation energy to the kinetic energy carried by cold baryons, and 
$\zeta_e$ is the ratio of the number of electrons to the number of baryons, taking into account the possibility 
of copious pair production via internal dissipation. The photospheric radius thus obtained would generally be above the typical radius of a Wolf-Rayet star.  Also, especially when the late jet is baryon-rich compared to that for prompt emission, the photospheric radius is likely to be located above the coasting radius. 
The comoving temperature at the photospheric radius is 
\begin{equation}
k T \simeq 30~{\rm eV}~L_{\rm LP, 48}^{1/4} r_{\rm ph, 13.5}^{-1/2} {(\Gamma_{\rm LP}/5)}^{-1/2},  
\end{equation}
and the observed typical energy is 
\begin{equation}
E^b \simeq 0.48~{\rm keV}~L_{\rm LP,48}^{5/12} r_{\rm ph,13.5}^{-2/3} {\left( \frac{\zeta_e/\epsilon_e}{10} \right)}^{1/6} {(1+z)}^{-1}.
\end{equation}
The observed X-ray emission shows a hard spectrum in the X-ray band, $F_{\rm LP} \propto E^{-1}$, which requires 
some process such as Comptonization by nonthermal electrons produced via internal dissipation around the photosphere. In this scenario, the variability timescale would be relatively short, $\delta T_{\rm var} \sim 21~{\rm s}~r_{i,13.5} {(\Gamma_{\rm LP}/5)}^{-2} (1+z)$, although dissipation itself may last for a longer time. 
 
As mentioned before, flares would also be caused by activities of the long-lasting central engine. 
However, it might not be easy to explain flares in this scenario. Flares seem to be caused by occasional larger dissipation of relativistic outflows but those with Lorentz factors whose values are larger than the values of prompt emission but may be smaller than the values of late prompt emission, $\Gamma_{\rm flare} \sim 10-50$~\cite{JFW10}. On the other hand, light curves of flares often show the exponential decay after the peak, which suggests relatively large emission radii of $r_i \sim 3 \times {10}^{14}~{\rm cm}~\Gamma_{\rm flare,1}^2 {\Delta T}_{\rm flare,2} {(1+z)}^{-1}$, if the decay of pulses is attributed to high-latitude emission. The typical radii seem above the photospheric radius, unless the jet is largely pair dominated. 
Here, one should keep in mind that the photospheric scenario and other scenarios are not mutually exclusive. 
For example, internal shocks may occur well above the photospheric radius as well as around the photospheric radius~\cite{RM05}. Magnetic reconnection is also one of the possibilities.  

\subsection{Magnetic Dissipation Scenario}
Relativistic jets launched by the central engine may be initially Poynting-dominated. If the outflow is still Poynting-dominated at the emission radii, without significant conversion into the kinetic energy, magnetic dissipation rather than shock dissipation of the bulk kinetic energy may lead to production of nonthermal particles. Although detailed scenarios for this are still unavailable due to lack of our knowledge on mechanisms of magnetic dissipation and associated particle acceleration, prompt and/or late prompt emission may be produced by internal dissipation of a significant fraction of the magnetic energy in the outflow. For example, Lyutikov (2006) argued that magnetic dissipation may occur around the radius where the MHD approximation breaks down, if the outflow is extremely magnetized. On the other hand, magnetic fields are distorted by internal shocks at $r_i \sim {10}^{13-15}$~cm, which may eventually lead to efficient magnetic 
reconnection~\cite{ZY11,MU10}. 

In this work, just for demonstrative purposes, we apply the jets-in-a-jet model~\cite{Gia+09} for the late prompt emission. In this scenario, the magnetic reconnection in a jet leads to many mini-blobs with relative Lorentz factors of $\sim \sqrt{\sigma}$. Indeed, radiation from such mini-blobs can reproduce highly variable light curves of prompt emission, though there remain potential problems~\citep[see, e.g.,][]{LNP09}.  
Assuming that the late jet has $\Gamma_{\rm LP} \sim 5$ with the magnetization parameter of $\sigma \sim 30$, the dissipation radius is estimated as $r_{i} \simeq 3.8 \times {10}^{15}~{\rm cm}~{(\Gamma_{\rm LP}/5)}^2 \sigma_{1.4} {\delta T}_{\rm var,2} {(1+z)}^{-1}$. 

The typical electron Lorentz factor is estimated as 
\begin{equation}
\gamma_e \sim \epsilon_e \sqrt{\sigma} \frac{m_p}{m_e} \simeq 0.92 \times {10}^{3} \sigma_{1.4}^{1/2} \epsilon_{e,-1}.
\end{equation}
The magnetic field in the downstream blob can be $B \sim 5.2 \times {10}^{2}~{\rm G}~L_{B,48}^{1/2} 
r_{i, 15.5}^{-1} {(\Gamma_{\rm LP}/5)}^{-1}$ (note that it does not have to be Poynting dominated since a significant fraction of the magnetic energy is dissipated there). 
Then, the typical synchrotron peak energy is estimated as\footnote{If one applies this model to GRB prompt emission, we have $E^b \simeq  0.57~{\rm MeV}~\epsilon_{e,-1}^2 \sigma_{2.5}^{3/2} L_{B,52}^{1/2} r_{i,15.5}^{-1} {(1+z)}^{-1}$.} 
\begin{equation}
E^b \simeq 0.13~{\rm keV}~\epsilon_{e,-1}^2 \sigma_{1.4}^{3/2} L_{B,48}^{1/2} r_{i,15.5}^{-1} {(1+z)}^{-1},
\end{equation} 
which seems consistent with observations. A hard spectrum, $F_{\rm LP} \propto E^{-1}$, may be attributed to synchrotron emission from nonthermal electrons accelerated at shocks caused by mini-blobs. Note that electrons are typically in the fast cooling regime, since $\gamma_{e,c} \simeq 0.14 L_{B,48}^{-1} r_{i,15.5} {(\Gamma_{\rm LP}/5)}^{3}$ (of course, the actual electron Lorentz factor should be larger than unity). Here, the lower bulk Lorentz factor with the lower magnetization is assumed for the late jet compared to the case of prompt emission, but it may not be the case in some models such as the fast rotating magnetar model~\cite{Met+10}. 
In order to have the synchrotron peak of $\sim 0.1$~keV, not only large radii but also low values of $\epsilon_e$ might be required in this scenario. For example, if $\sigma \sim {10}^{5}$ and $r_i \sim {10}^{17}$~cm, $\epsilon_e \sim {10}^{-3}$ and the short variability time are expected, depending on the scenario.

\section[]{Associated High-Energy Emission}
Next, we consider consequences of the late internal dissipation model for high-energy emission. One can consider two possibilities of high-energy emission. 
One is EIC emission produced by electrons accelerated at the external shock caused by the early jet, which is responsible for the prompt emission and the observed standard afterglow component. Late prompt photons from inner radii are naturally upscattered in the late internal dissipation model, and predictions are not sensitive to details of late internal dissipation models. In this paper, we especially discuss this possibility in detail (see the next section). 
The other is the high-energy emission from the emission radius at which internal dissipation occurs, e.g., SSC emission from the late jet. Obviously, predictions of high-energy emission depend on each scenario. This possibility is also discussed in this section. 

\subsection{High-Energy Afterglow Emission}
First, we discuss EIC emission caused by interactions with late prompt photons and electrons accelerated at the external shock of the early jet producing prompt emission. As demonstrated below, this EIC emission is useful as a test of the late internal dissipation model. We here give analytical considerations, but more detailed results with numerical calculations are provided in the next section.
 
We can think that one of the two components is the standard afterglow component from the early jet. For an adiabatic relativistic blast wave expanding into the interstellar medium (ISM)~\cite{BM76}, we obtain the bulk 
Lorentz factor as 
\begin{equation}
\Gamma (T) \simeq 44~\mathcal{E}_{k,53}^{1/8} n_0^{-1/8} T_{3}^{-3/8} {(1+z)}^{3/8},
\end{equation}
and the external shock radius is estimated as 
\begin{equation}
R(T) \simeq 2.3 \times {10}^{17}~{\rm cm}~ \mathcal{E}_{k,53}^{1/4} n_0^{-1/4} T_{3}^{1/4} {(1+z)}^{-1/4}, 
\end{equation}
where $\mathcal{E}_{k}$ is the isotropic kinetic energy of the ejecta and $n$ is the ISM density. 

Electrons would be accelerated at the external FS. The injection Lorentz factor of electrons is estimated as
\begin{equation}
\gamma_{e,m} \simeq 2.3 \times {10}^{2} \epsilon_{ef,-2} f_{ef}^{-1} (g_p/g_{2.4}) \mathcal{E}_{k,53}^{1/8} n_0^{-1/8} T_3^{-3/8} {(1+z)}^{3/8},
 \end{equation}
where $g_p=(p-1)/(p-2)$ and $p$ is the spectral index of FS electrons. Here $\epsilon_{ef}$ is the fraction of the internal energy of the shocked ISM transferred to non-thermal electrons at the external FS, and $f_{ef}$ is the number fraction of electrons injected to the acceleration process at the external FS~\cite{EW05}. The cooling Lorentz factor of electrons is estimated by $t_{\rm dyn}=t_{\rm cool}$, and we have
\begin{equation}
\gamma_{e,c} \simeq 6.2 \times {10}^{4} \epsilon_{Bf,-3}^{-1} \mathcal{E}_{k,53}^{-3/8} n_0^{-5/8} T_3^{1/8} {(1+z)}^{-1/8} {(1+Y)}^{-1}
 \end{equation}
where $t_{\rm dyn}=\tilde{\Delta}/c \approx (4/\kappa) \Gamma c T$ is the dynamical timescale, $t_{\rm cool}$ is the electron cooling timescale, and $Y$ is the total Compton $Y$ parameter. Here, $\kappa$ is set to 4 in this work~\cite{PK04}, and $\epsilon_{Bf}$ is the fraction of the internal energy of the shocked ISM transferred to the downstream magnetic field. 
In the slow cooling case ($\gamma_{e,m} < \gamma_{e,c}$) with a constant $Y$, the steady electron distribution 
is $d \mathcal{N}_e/d \gamma_e \propto \gamma_e^{-p}$ for $\gamma_{e,m} \leq \gamma_e < \gamma_{e,c}$ and 
$d \mathcal{N}_e/d \gamma_e \propto \gamma_e^{-p-1}$ for $\gamma_{e,c} \leq \gamma_e$. In the fast cooling 
case ($\gamma_{e,c} < \gamma_{e,m}$) with a constant $Y$, the steady electron distribution is 
$d \mathcal{N}_e/d \gamma_e \propto \gamma_e^{-2}$ for $\gamma_{e,c} \leq \gamma_e < \gamma_{e,m}$ and 
$d \mathcal{N}_e/d \gamma_e \propto \gamma_e^{-p-1}$ for $\gamma_{e,m} \leq \gamma_e$.

These electrons upscatter late prompt photons at the vicinity of the FS. The expected EIC luminosity is very roughly written as $L_{\rm EIC} \sim {\rm min} (\mathcal{Y}_{\rm EIC} L_{\rm LP}, L_e)$~\citep[e.g.,][]{Fan+08}, where $\mathcal{Y}_{\rm EIC}$ is introduced as the ratio of the EIC energy flux to the seed photon energy flux. In the slow cooling case, noting that $F_{\rm EIC} (E) \sim  \int d \gamma_e \frac{d \tau_{e}}{d \gamma_e} F_{\rm LP} (\gamma_e,E)$, where $\gamma_e \frac{d \tau_e}{d \gamma_e}  \sim \tau_T {(\gamma_e/\gamma_{e,m})}^{-p+1}$ for $\gamma_{e,m} \leq \gamma_e<\gamma_{e,c}$ and $\gamma_e \frac{d \tau_e}{d \gamma_e} \sim \tau_T  {(\gamma_{e,c}/\gamma_{e,m})}^{-p+1} {(\gamma_e/\gamma_{e,c})}^{-p}$ for $\gamma_e \geq \gamma_{e,c}$, 
the resulting EIC spectrum in the Thomson limit is expressed as
\begin{equation}
E F_{\rm EIC} (E) 
\propto
\left\{ \begin{array}{ll} 
E^{2-\beta_l}
& \mbox{($E < E_{\rm EIC}^m$)}\\
E^{(3-p)/2}
& \mbox{($E_{\rm EIC}^m \leq E < E_{\rm EIC}^c$)}\\
E^{(2-p)/2}
& \mbox{($E_{\rm EIC}^c \leq E$)}
\end{array} \right.
\end{equation}
where $\tau_T \sim (\sigma_T \mathcal{N}_e/4 \pi R^2)$ is the Thomson optical depth, $\mathcal{N}_e$ is the number of electrons, $\beta_l$ is the low-energy photon index of late prompt emission, and   
\begin{eqnarray}
E_{\rm EIC}^m &\approx& \gamma_{e,m}^2 E^b \nonumber \\
&\simeq& 5.2~{\rm MeV}~E_{0.1~\rm keV}^b\epsilon_{ef,-2}^2  f_{ef}^{-2} {\left(\frac{g_p}{g_{2.4}} \right)}^{2} \mathcal{E}_{k,53}^{1/4} n_0^{-1/4} {\left(\frac{T_3}{1+z} \right)}^{-3/4}\\
E_{\rm EIC}^c &\approx& \gamma_{e,c}^2 E^b \nonumber \\ 
&\simeq& 95~{\rm GeV}~E_{0.1~\rm keV}^b \epsilon_{Bf,-3}^{-2} \mathcal{E}_{k,53}^{-3/4} n_0^{-5/4} {\left(\frac{T_3}{1+z} \right)}^{1/4} {\left(\frac{2}{1+Y} \right)}^2  
\end{eqnarray}
The contribution below $E_{\rm SSC}^m$ mainly comes from interactions between electrons with $\sim \gamma_{e,m}$ 
and photons with $E<E^b$, while the contribution in the range  $E_{\rm SSC}^m \leq E < E_{\rm SSC}^c$ comes from interactions between electrons with $\gamma_{e,m} < \gamma_e \leq \gamma_{e,c}$ and photons with $\sim E^b$. 
The EIC flux at $E_{\rm EIC}^c$ is also estimated from $E_{\rm EIC}^c F_{\rm EIC}^c \approx x Y_{\rm EIC} 
(E^c F^c)$, where the EIC Compton $Y$ parameter, $Y_{\rm EIC}$, is introduced as the ratio of 
the EIC energy loss rate to the synchrotron energy loss rate (and it is different from $\mathcal{Y}_{\rm EIC}$). Here, $x \lesssim 1$ is a factor coming from the fact that the EIC emission is anisotropic. 
 
The Thomson limit has been consider so far. 
However, the Klein-Nishina (KN) effect often becomes important at sufficiently high energies~\citep[e.g.,][]{GG03}. 
In our case, there are three characteristic energies.
\begin{eqnarray}
E_{\rm KN}^m &\approx& \Gamma \gamma_{e,m} m_e c^2/(1+z) \\
E_{\rm KN}^c &\approx& \Gamma \gamma_{e,c} m_e c^2/(1+z) \\
E_{\rm KN}^b &\approx& 2 \Gamma^2 m_e^2 c^4/E^b/(1+z)^2 
\end{eqnarray}
When the KN effect becomes important, the EIC spectrum has breaks. Here, let us introduce $E_{\rm KN,1}$ as the first break energy due to the KN effect.  
When the KN break exists above the EIC peak, instead of Equation~(15), we have 
\begin{equation}
E F_{\rm EIC} (E) 
\propto
\left\{ \begin{array}{ll} 
E^{2-\beta_l}
& \mbox{($E < E_{\rm EIC}^m$)}\\
E^{(3-p)/2}
& \mbox{($E_{\rm EIC}^m \leq E < E_{\rm EIC}^c$)}\\
E^{(2-p)/2}
& \mbox{($E_{\rm EIC}^c \leq E < E_{\rm KN,1}$)}\\
E^{\beta_l-p}
& \mbox{($E_{\rm KN,1} \leq E$)}
\end{array} \right.
\end{equation}
Here 
\begin{equation}
E_{\rm KN,1}=E_{\rm KN}^b \simeq 9.9~{\rm TeV}~{(E_{0.1~\rm keV}^b)}^{-1} \mathcal{E}_{k,53}^{1/4} n_0^{-1/4} T_{3}^{-3/4} {(1+z)}^{-5/4}.
\end{equation}
This case is typical for our adopted parameters, and the EIC emission at $E > E_{\rm KN,1}$ is dominated by radiation from electrons with $\gamma_e \sim E/\Gamma m_e c^2 (1+z)$ interacting with seed photons with the energy of $\sim \Gamma^2 m_e^2 c^4/E{(1+z)}^2$ via the Thomson scattering. 

If the KN break appears below $E_{\rm EIC}^c$, we obtain  
 \begin{equation}
E F_{\rm EIC} (E) 
\propto
\left\{ \begin{array}{ll}  
E^{2-\beta_l}
& \mbox{($E < E_{\rm EIC}^m$)}\\
E^{(3-p)/2}
& \mbox{($E_{\rm EIC}^m \leq E < E_{\rm KN,1}$)}\\
E^{\beta_l-p}
& \mbox{($E_{\rm KN,1} \leq E$)}
\end{array} \right.
\end{equation}
where
\begin{equation}
E_{\rm KN,1}=E_{\rm KN}^c \simeq 690~{\rm GeV}~\epsilon_{Bf,-3}^{-1} \mathcal{E}_{k,53}^{-1/4} n_0^{-3/4} T_{3}^{-1/4} {(1+z)}^{-3/4} \frac{2}{1+Y}.
\end{equation} 
If $\gamma_{e,m}$ and/or $E^b$ are too large, one expects the deep KN regime. In this case, we have
\begin{equation}
E F_{\rm EIC} (E) 
\propto
\left\{ \begin{array}{ll} 
E^{2-\beta_l}
& \mbox{($E < E_{\rm KN,1}$)}\\
E^{\beta_l-p+1} 
& \mbox{($E_{\rm KN,1} \leq E < E_{\rm KN,2}$)}\\
E^{\beta_l-p} 
& \mbox{($E_{\rm KN,2} \leq E$)}
\end{array} \right.
\end{equation}
Here, 
\begin{equation}
E_{\rm KN,1}=E_{\rm KN}^m \simeq 5.1~{\rm GeV}~\epsilon_{ef,-2} f_{ef}^{-1} (g_p/g_{2.4}) \mathcal{E}_{k,53}^{1/4} n_0^{-1/4} T_{3}^{-3/4} {(1+z)}^{-1/4} 
\end{equation}  
and $E_{\rm KN,2} \equiv \Gamma \gamma_{e,c} m_e c^2/{(1+z)}^2$ is the second KN break.
This EIC spectrum is anticipated in the prior emission model for shallow-decay emission~\cite{Mur+10}.

We are interested especially in cases where the EIC flux exceeds the afterglow SSC flux. For this purpose, we next estimate the SSC flux. The SSC emission has been studied by many authors~\citep[see reviews, e.g.,][]{FP08}, so that we here discuss it just briefly. 
The characteristic energies of the SSC emission are obtained as~\citep[e.g.,][]{SE01,ZM01b}
\begin{eqnarray}
E_{\rm SSC}^m &\simeq& 4.7~{\rm keV}~{(g_p/g_{2.4})}^4 f_{ef}^{-4}
{\epsilon}_{ef,-2}^4 {\epsilon}_{Bf,-3}^{1/2} \nonumber \\ 
&\times& \mathcal{E}_{k,53}^{3/4} n_0^{-1/4}
T_3^{-9/4} {(1+z)}^{5/4} \\
E_{\rm SSC}^c &\simeq& 490~{\rm GeV}
~{\left( \frac{1+Y}{2} \right)}^{-4} {\epsilon}_{Bf,-3}^{-7/2} \nonumber \\ 
&\times& {\mathcal{E}}_{k,53}^{-5/4} n_0^{-9/4} {T}_{3}^{-1/4} {(1+z)}^{-3/4}.
\end{eqnarray}
For the slow cooling case that we are interested in, the SSC spectrum in the Thomson limit is expressed as
\begin{equation}
E F_{\rm SSC} (E) 
\propto
\left\{ \begin{array}{ll} 
E^{4/3}
& \mbox{($E < E_{\rm SSC}^m$)}\\
E^{(3-p)/2}
& \mbox{($E_{\rm SSC}^m \leq E < E_{\rm SSC}^c$)}\\
E^{(2-p)/2}
& \mbox{($E_{\rm SSC}^c \leq E$)}
\end{array} \right.
\end{equation} 
Note that only the first SSC component is important, since the second SSC component is typically negligible due to the KN suppression. The energy flux at the SSC peak is also evaluated as
\begin{eqnarray}
E_{\rm SSC}^c~F_{\rm SSC}^{c}~&\simeq&~1.2~\times~{10}^{-7}
~{\rm GeV}~{\rm cm}^{-2}~{\rm s}^{-1}~
Y_{\rm SSC} {\left( \frac{1+Y}{2} \right)}^{p-3}
d_{L,27.5}^{-2}
\nonumber \\
&\times&
 {\left( \frac{g_p}{g_{2.4}} \right)}^{p-1}  f_{ef}^{2-p} \epsilon_{ef,-2}^{p-1} 
\epsilon_{Bf,-3}^{p-2} \mathcal{E}_{k,53}^{\frac{p}{2}} n_0^{\frac{p-2}{2}}
{\left(\frac{T_3}{1+z} \right)}^{-\frac{p}{2}},
\end{eqnarray}
by which we can normalize the SSC spectrum. As a result, the EIC flux and SSC flux are roughly related as $E_{\rm EIC}^c F_{\rm EIC}^c/E_{\rm SSC}^c F_{\rm SSC}^c \sim x Y_{\rm EIC}/Y_{\rm SSC}$. See Section 4 for detailed discussions on the relative importance of each component.

The KN effect may become important in the cases we consider here. For our typical parameters, the KN break is located above $E_{\rm SSC}^c$, which is
\begin{eqnarray}
E_{\rm KN} &\approx& \frac{\Gamma^2}{({1+z)}^2} \frac{m_e^2 c^4}{E^c} \nonumber \\
&\simeq& 1.4~{\rm TeV}~\epsilon_{Bf,-3}^{3/2} \mathcal{E}_{k,53}^{5/8} n_0^{7/8} T_3^{1/8} {(1+z)}^{-9/8} {\left(\frac{1+Y}{2} \right)}^{2}.
\end{eqnarray}
In general, SSC spectra can be complicated and consist of several breaks~\cite{NAS09,Wan+10}. Hence, we numerically calculate the SSC emission taking into account the KN effect. 

We can also calculate light curves, once the dynamical evolution of the blast wave is given. 
In the next section, we show the resulting light curves of the EIC and SSC emission. 
Note that the temporal behavior would change after the jet break time of $T_{j} \sim {10}^{5}$~s \cite{Rho99,SPH99} (where $\Gamma \theta_j \sim 1$), but throughout this work we focus on the behavior before $t_j$.

\subsection{High-Energy Late Prompt Emission}
High-energy emission is expected from the late jet itself, as mentioned before. For example, one can expect SSC emission as well as synchrotron emission if electrons are accelerated in the magnetized region. If protons are also accelerated up to very high energies, hadronic gamma rays are produced via photomeson and photopair production, and proton-synchrotron radiation. 
Predictions depend on models, which are quite uncertain. Hence, in this subsection, we just provide analytical considerations on several interesting cases.  

In the photospheric scenario, the injection Lorentz factor of electrons should be $\gamma_{e,m}\sim 1$ to produce the hard spectral component of $F_{\rm LP} \propto E^{-1}$ by Comptonization~\citep[e.g.,][]{Tho94,Iok+07}. The IC spectrum may be extended up to high energies, but high-energy photons cannot avoid attenuation by pair-production. Murase \& Ioka (2008) showed that the pair-production break (or cutoff) should be around $E_{\rm cut} \approx \frac{480}{11} \Gamma_{\rm LP} m_e c^2/(1+z)$ in the pair-photospheric scenario (for $\beta_h=2$), which suggests that high-energy emission above GeV is not expected in the one-zone case. 
In the multi-zone case, relativistic electrons may be produced at outer radii. For example, internal dissipation 
may occur above the photospheric radius, leading to EIC emission with the typical energy of $\sim 
\gamma_{e,m}^2 E^b \sim {\rm GeV}$. However, we will not discuss here such more complicated possibilities.  

In the magnetic dissipation or late-internal shock scenarios, the typical emission radii are much larger, so that it is easier to expect high-energy gamma rays that escape from the source.  Here, as a demonstrative example, we consider the SSC emission in the jets-in-a-jet model described in the previous section, which is sufficient for our purposes in this work.  First, in the Thomson limit, the typical SSC energy is estimated as 
\begin{equation}
E_{\rm SSC}^{b} = 2 \gamma_{e,m}^2 E^b \simeq 210~{\rm MeV}~\epsilon_{e,-1}^4 \sigma_{1.4}^{5/2} L_{B,48}^{1/2} r_{i,15.5}^{-1} {(1+z)}^{-1}.
\end{equation}
Introducing the Compton $Y$ parameter, $Y_{\rm LP}$, the SSC flux at $E_{\rm SSC}^b$ is written as
\begin{eqnarray}
E_{\rm SSC}^b F_{\rm SSC}^b &\approx& Y_{\rm LP} E^b F_{\rm LP}^b \nonumber \\
&\sim& 2.5 \times {10}^{-6}~{\rm GeV}~{\rm cm}^{-2} {\rm s}^{-1}~\frac{Y_{\rm LP} L_{e,48}}{(1+Y_{\rm LP})} \frac{1+z}{d_{L,27.5}^{2}}.
\end{eqnarray}
In the Thomson limit, $Y_{\rm LP}$ can be approximated as $Y_{\rm LP} \approx \frac{-1+\sqrt{1+4\epsilon_e/\epsilon_B}}{2}$.
But, the KN effect may actually become important at sufficiently high-energies. When the KN break exists above $E_{\rm SSC}^b$, we have 
\begin{equation}
E F_{\rm SSC} (E) 
\propto
\left\{ \begin{array}{ll} 
E^{2-\beta_l}
& \mbox{($E < E_{\rm SSC}^b$)}\\
E^{2-\beta_h}
& \mbox{($E_{\rm SSC}^b \leq E < E_{\rm KN,1}$)}\\
E^{\beta_l+2-2\beta_h}
& \mbox{($E_{\rm KN,1} \leq E$)}
\end{array} \right. \label{LPSSC}
\end{equation}
The KN break is given by $E_{\rm KN,1} = {(\Gamma_{\rm em} m_e c^2/(1+z)E^b)}^2 E^b \simeq 1.3~{\rm TeV}~{(\Gamma_{\rm LP}/5)}^2 \epsilon_{e,-1}^{-2} \sigma_{1.4}^{-1/2} L_{B,48}^{-1/2} r_{i,15.5} {(1+z)}^{-1}$.   
If $E_{\rm KN,1} < E_{\rm SSC}^b$, the spectrum is in the deep KN regime, and we obtain 
\begin{equation}
E F_{\rm SSC} (E) 
\propto
\left\{ \begin{array}{ll} 
E^{2-\beta_l}
& \mbox{($E < E_{\rm KN,1}$)}\\
E^{\beta_l+2-2\beta_h}
& \mbox{($E_{\rm KN,1} \leq E$)}
\end{array} \right.
\end{equation}
where $E_{\rm KN,1} = \Gamma_{\rm em} \gamma_{e,m} m_e c^2/(1+z) \simeq 12~{\rm GeV}~\epsilon_{e,-1}(\Gamma_{\rm LP}/5) \sigma_{1.4} {(1+z)}^{-1}$. 
In the fast cooling case, the resulting spectra can be more complicated especially when $Y_{\rm LP}$ in the Thomson limit is so large that the distribution of electrons is affected by the KN effect~\citep[see, e.g.,][]{NAS09,BDD09}. But the above expressions are reasonable for moderately small values of $Y_{\rm LP}$.    

In the late internal dissipation scenario, the pair-creation process is crucial for high-energy gamma-ray emission. The optical depth for pair production is estimated as~\citep[e.g.,][]{LS01,MI08,GZ08} 
\begin{equation}
\tau_{\gamma \gamma} \approx 0.1 \sigma_T l \frac{L_{\rm LP}^b}{4 \pi r_i^2 \Gamma_{\rm em} c (1+z)E^b} {\left( \frac{{(1+z)}^2EE^b}{\Gamma_{\rm em}^2 m_e^2 c^4} \right)}^{\beta-1},
\end{equation}
where $l$ is the comoving width. Assuming $l \sim r_i/\Gamma_{\rm em}$, the pair-production break (or cutoff) is estimated as $E_{\rm cut} \simeq 8.5~{\rm GeV}~{(L_{\rm LP,48}^b)}^{-\frac{1}{\beta-1}} r_{i,15.5}^{\frac{1}{\beta-1}} {(\Gamma_{\rm LP}/5)}^{\frac{2 \beta}{\beta-1}} \sigma_{1.4}^{\frac{\beta}{\beta-1}} {(E_{0.1~\rm keV}^b)}^{\frac{2\beta}{\beta-1}} {(1+z)}^{\frac{3-2\beta}{\beta-1}}$ in the magnetic dissipation model. 
At energies higher than this energy, the spectrum is suppressed or may have a cutoff.

In Figures~1 and 2, we show SSC spectra which are calculated analytically using Equation~(\ref{LPSSC}). 
The parameters are described in the caption of Figure~1. The pair-production opacity is taken into account by $1/(1+\tau_{\gamma \gamma})$~\cite{Bar06}.  
From Figures~1 and 2, we see that it is difficult to detect high-energy late prompt emission from distant bursts, but is possible for nearby bursts. The SSC peak is expected around the GeV range, which may be 
reached by \textit{Fermi} if GRBs occur at $z \lesssim 0.7$. Very high energy gamma rays above $\sim 30$~GeV 
are also detectable with the future CTA, although its detectablity depends on the pair-creation opacity both 
inside and outside the source.  In Figures~1 and 2, the KN break is seen around TeV but the attenuation by 
pair-creation masks it. The light curves of the high-energy gamma rays basically follow the observed X-ray light curve of the late prompt emission.  

\begin{figure}
\includegraphics[width=\linewidth]{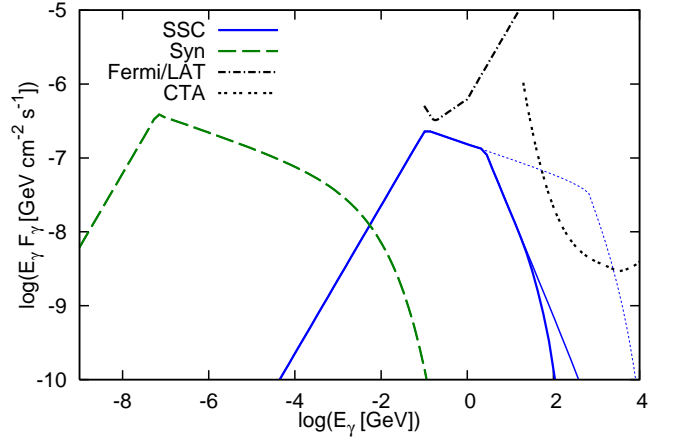}
\caption{Spectra of synchrotron and SSC emission in the magnetic dissipation model for late prompt emission at $T={10}^{3}$~s. The source redshift is taken to $z=1$. 
Assumed parameters are: $L_B|_{T_a^{\prime}} =L_e|_{T_a^{\prime}} = {10}^{48.5}~{\rm erg}~{\rm s}^{-1}$, $\sigma={10}^{1.4}$, $\Gamma_{\rm LP}=5$, and $r_{i}={10}^{15.75}$~cm. 
The thick solid curve represents an SSC spectrum taking into account attenuation by pair-creation both inside and outside the source. An SSC spectrum shown as the thin sold curve includes only the source attenuation, while the thin dotted curve spectrum does not include either of them.      
The \textit{Fermi}/LAT and CTA sensitivities (with the duty factor of 30~\%) are also overlayed (CTA Consortium 2010). The LAT sensitivity curves in the sky survey mode are used for the long time observations, although the possible continuous observations by LAT may improve the detectability by a factor of 3-5 (e.g., Gou \& M\'esz\'aros 2007).}
\end{figure}
\begin{figure}
\includegraphics[width=\linewidth]{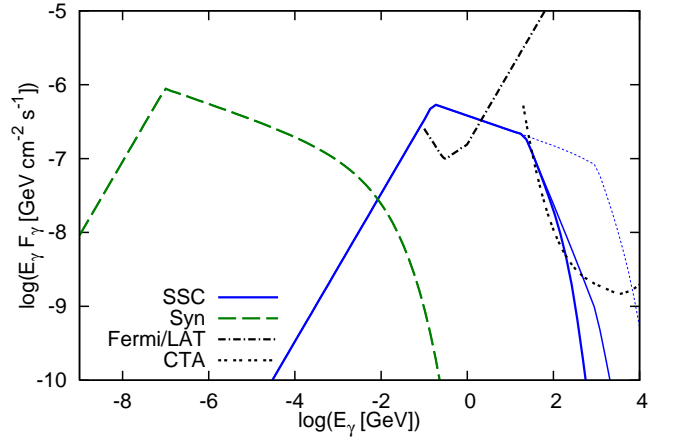}
\caption{Same as Figure~1, but $T={10}^{3.6}$~s and $z=0.3$.}
\end{figure}

One can also calculate the SSC emission in the late internal shock scenario similarly to how it was done in the previous paragraph, by changing parameters.
In this paragraph, we briefly discuss the hadronic emission, although detailed studies are beyond the scope of this work. In the late internal shock scenario, not only electron but also protons may be accelerated up to very high energies. Even in the magnetic dissipation scenario, protons may be accelerated~\cite{Gia10}, although a large baryon loading may not be expected. Hadronic emission in the late internal dissipation model was considered and discussed in Murase (2007). The Lorentz factor of the late jet might be relatively small, and the late jet might be more baryon-rich compared to the early jet making prompt emission. 
Then, as shown in Murase (2007) and Murase \& Nagataki (2006), copious soft photon fields in the late jet lead to a high meson production efficiency given by
\begin{equation}
f_{p \gamma} \sim 1.4 \frac{L_{\rm LP,48}^b}{r_{i,15.5} {(\Gamma_{\rm em}/25)}^2 (1+z) E^b_{0.1~\rm keV}}  {(E_p/E_p^b)}^{\beta-1},  
\end{equation}
where the multi-pion production effect (which is a factor of three) is taken into account. 
The expected neutrino flux is comparable to or maybe larger than that of prompt emission, since the meson production efficiency is high while the total radiation energy of flares or late prompt emission is $\sim 10$~\% of that of prompt emission~\cite{Fal+07,Chi+10} (and the kinetic or magnetic energy may be larger if the radiation efficiency is low). 
Hadronic gamma rays are also expected as well as neutrinos. The efficient pair-production in the source induces electromagnetic cascades. Assuming an $E^{-2}$ spectrum for cascade 
emission, the gamma-ray flux is crudely estimated as 
\begin{eqnarray}
E F_{p \gamma} &\sim& \frac{1+z}{4 \pi d_L^2} \frac{5}{8} \frac{L_{\rm CR}}{\mathcal{R}} \nonumber \\
&\sim& 1.6 \times {10}^{-6}~{\rm GeV}~{\rm cm}^{-2} {\rm s}^{-1}~L_{\rm CR,49} \frac{20}{\mathcal{R}} \frac{1+z}{d_{L,27.5}^{2}},
\end{eqnarray}
where $\mathcal{R}$ is the conversion factor from the total energy amount of protons into the energy amount of protons per energy decade, $\mathcal{R} \sim 20$ for $p=2$.  Although the detailed calculation is beyond the scope of this work, this suggests the potential importance of hadronic gamma rays for nearby GRBs. 

Although predictions for the high-energy late prompt emission are model-dependent, once high-energy gamma rays are detected in sufficient amounts, they would be useful to distinguish between the various uncertain mechanisms of late prompt emission.

\section{Numerical Results of High-Energy Afterglow Emission}
In the previous section, we gave analytical estimates of EIC and SSC emission produced by relativistic 
electrons accelerated at the external FS. As we discussed, the EIC emission does not depend on details of late internal dissipation mechanisms, and is useful as a good probe of the different scenarios.
In this section we calculate numerically the EIC emission, which provides significantly more accurate results than the analytical estimates. This is because: (1) the EIC emission is anisotropic, which leads to suppression by a factor of $x$; (2) the KN suppression becomes important at high energies above $E_{\rm KN,1}$; (3) the influence on the electron distribution is complicated if the EIC/SSC cooling is efficient and the KN effect is relevant. 

In order to calculate the EIC emission, we need to consider the equal-arrival-time surface of upscattered photons. The expression for the EIC emission is written as (see Appendix A) 
\begin{eqnarray}
F_{\rm EIC} (T)=\frac{3}{2} \sigma_T \int \frac{d r}{r} (1-\cos \tilde{\theta}) \int d \gamma_e \, \frac{d n_e}{d \gamma_e} \tilde{\Delta} \int d y \, (1-\xi) \nonumber \\ 
\times \left[ 1- 2 y +2 y^2 + \frac{\xi^2}{2(1-\xi)} \right]
\frac{F_{\rm LP} (r) G(\varepsilon)}{{(1+\Gamma^2 \theta^2)}^2} \label{EICformula}
\end{eqnarray}
where 
$y \equiv \frac{\xi m_e c^2}{2(1-\cos \tilde{\theta}) \gamma_e \varepsilon (1-\xi)}$ 
and 
$\xi \equiv \frac{(1+z) (1+ \Gamma^2 \theta^2) E}{2 \Gamma \gamma_e m_e c^2}$.
The scattering angles $\theta$ and $\tilde{\theta}$ of EIC photons are measured in the central engine frame and the comoving frame, respectively. 
The function $G(\varepsilon)$ represents the spectral shape of seed photons with energy $\varepsilon$ in the comoving frame (e.g., $\varepsilon^b = (1+z) E^b/2 \Gamma$).  
In the case of a broken power-law spectrum for late prompt emission, it is 
$G(\varepsilon)={(\varepsilon/\varepsilon^b)}^{-\beta_l+1}$ for $\varepsilon < \varepsilon^b$ and  
$G(\varepsilon)={(\varepsilon/\varepsilon^b)}^{-\beta_h+1}$ for $\varepsilon^b \leq \varepsilon$, respectively. 

The input parameters required for the calculations are basically determined by afterglow observations at X-ray and optical bands. We set typical parameters following Ghisellini et al. (2009).  As for the electron distribution, we exploit the standard external FS model \citep[e.g.,][]{MR97,SPN98} and adopt the 
following fiducial parameter set: 
$\mathcal{E}_k={10}^{53.5}$~erg, $n=1~{\rm cm}^{-3}$, $\epsilon_{ef}={10}^{-2}$, $\epsilon_{Bf}={10}^{-3}$ and $p=2.4$. 

For the late prompt emission, the seed photon spectrum is assumed to be a broken power-law spectrum with 
$\beta_l = 1$ for $E<E^b$ and $\beta_h = 2.2$ for $E^b \leq E$, with ${E'}^b={10}^{2}-{10}^{2.5}$~eV. The break time is set to $T_a^{\prime}={10}^3$~s and the late prompt luminosity at $T_a^{\prime}$ is taken as $L_{\rm LP}^b|_{T_a^{\prime}}={10}^{48}-{10}^{48.5}~{\rm erg}~{\rm s}^{-1}$ (which means $\mathcal{E}_{\rm LP,X}/\epsilon_{ef} \mathcal{E}_k \sim 1$). The temporal indices are set to $\alpha_{\rm fl}=0.2$ and $\alpha_{\rm st}=1.5$.
Also, assuming $r_i={10}^{13.5}$~cm, the high-energy cutoff due to pair creation is determined from $E_{\rm cut} \simeq 8.5~{\rm GeV}~{(L_{\rm LP,48}^b)}^{-\frac{1}{\beta-1}} r_{i,15.5}^{\frac{1}{\beta-1}} {(\Gamma_{\rm em}/25)}^{\frac{2 \beta}{\beta-1}} {(E_{0.1~\rm keV}^b)}^{\frac{2\beta}{\beta-1}} {(1+z)}^{\frac{3-2\beta}{\beta-1}}$~\citep[e.g.,][]{GZ08,MI08} with the attenuation factor of $1/[1+\tau_{\gamma \gamma} (E)]$~\cite{Bar06}, and the low-energy cutoff due to synchrotron self-absorption is given from the blackbody limit~\citep[e.g.,][]{SZ09}. But, note that those cutoff energies are not relevant for our results. 
Jet opening angles of both the jets are set to $\theta_j=0.2$ and their bulk Lorentz factors are assumed to be larger than $1/\theta_j$.  

In this section, we discuss the results on high-energy afterglow emission, that is, EIC and SSC components from the early jet. One should keep in mind that SSC emission from the late jet, which was discussed in the previous section, may also exist. In the late internal shock and magnetic dissipation scenarios, one could expect $\sim 1-10$~GeV gamma rays via the SSC mechanism, which are potentially important for \textit{Fermi} if nearby and/or energetic GRBs occur. But its predictions are highly model dependent, and very high energy emission is not expected when $r_i$ and/or $\Gamma_{\rm em}$ are small enough (e.g., in the dissipative photosphere scenario), so that it will not be shown here. 
On the other hand, the EIC emission considered here is independent of various late internal dissipation scenarios. Even if the SSC emission from the late jet exists, this EIC and/or SSC components from the early jet will typically be dominant at very high energies. Therefore, our results on the EIC emission provide the most conservative high-energy predictions of the late internal dissipation model.

\begin{figure}
\includegraphics[width=\linewidth]{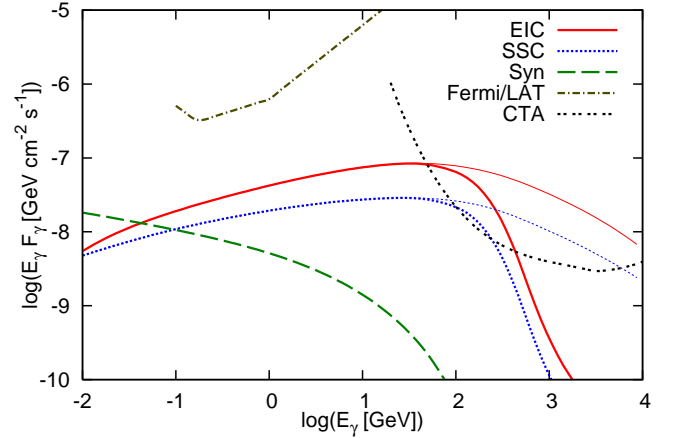}
\caption{Gamma-ray spectra of EIC emission in the late internal dissipation model for GRB afterglows, caused by Compton scatterings of X-ray photons by electrons accelerated at the external shock. Calculation are numerically performed according to Equation~(\ref{EICformula}), taking into account the equal-arrival-time surface. The observation time is set to $T={10}^{3}$~s and the source redshift is taken as $z=0.3$. 
Relevant parameters for the late jet are $L_{\rm LP}^b|_{T_a^{\prime}} = {10}^{48.5}~{\rm erg}~{\rm s}^{-1}$, $T_a^{\prime}={10}^{3}$~s, ${E'}^b=0.1$~keV, $\alpha_{\rm fl}=0.2$, and $\alpha_{\rm st}=1.5$.
Relevant parameters for the standard afterglow component are $\mathcal{E}_{k}= {10}^{53.5}~{\rm erg}$, $\epsilon_{ef}={10}^{-2}$, $\epsilon_{Bf}={10}^{-3}$, $n=1~{\rm cm}^{-3}$, and $p=2.4$.
For comparison, we also show the assumed synchrotron spectrum and the resulting SSC spectrum.   
Thick curves represent cases where the EBL attenuation is taken into account, while thin ones do not. Note that the attenuation by pair creation in the source is considered.  
The {\em Fermi}/LAT and CTA sensitivities (with the duty factor of 30~\%) are also overlayed (CTA Consortium 2010). 
}
\end{figure}
\begin{figure}
\includegraphics[width=\linewidth]{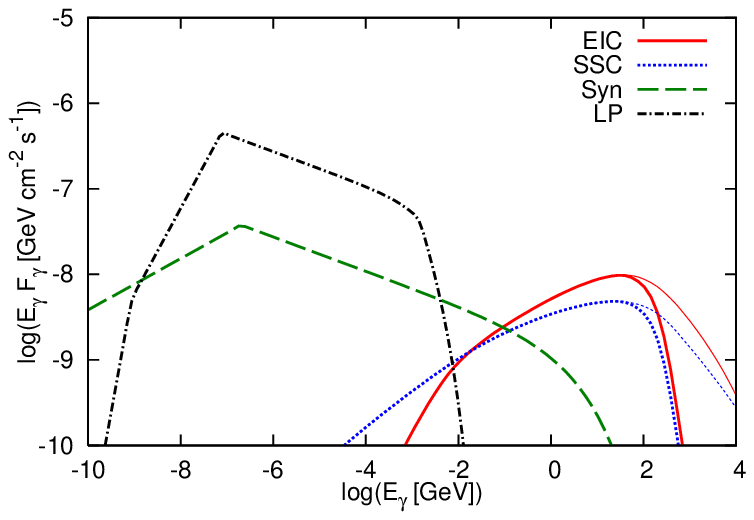}
\caption{Spectra of early and late jets in the late dissipation model, considered in this work. Syn and SSC come from synchrotron and SSC emission by relativistic electrons accelerated at the external shock of the early jet. LP represents the assumed seed photon spectrum from the late jet, which is responsible for shallow-decay X-ray emission, and EIC is the EIC emission by Compton scatterings of X-ray photons by electrons accelerated at the external shock.
The observation time is set to $T={10}^{3.6}$~s and the source redshift is taken as $z=0.3$. 
Here, relevant parameters for the late jet are $L_{\rm LP}^b|_{T_a^{\prime}} = {10}^{48}~{\rm erg}~{\rm s}^{-1}$, $T_a^{\prime}={10}^{3}$~s, ${E'}^b=0.1$~keV, $\alpha_{\rm fl}=0.2$, and $\alpha_{\rm st}=1.5$. Parameters for the standard afterglow component are the same as those used in the caption of Figure~3.
Thick curves represent cases where the EBL attenuation is taken into account, while thin ones do not. The attenuation by pair creation in the source is also considered.}
\end{figure}
\begin{figure}
\includegraphics[width=\linewidth]{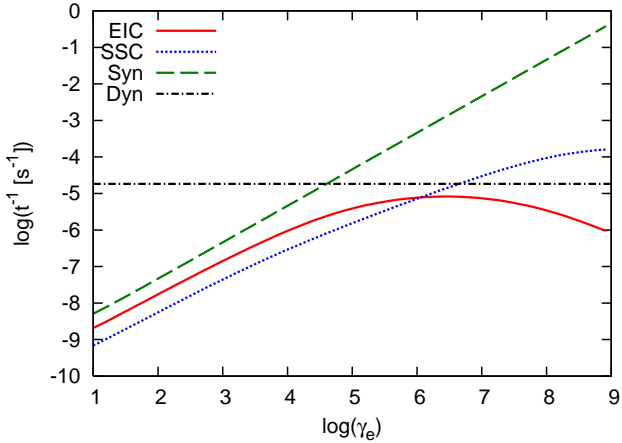}
\caption{Electron cooling timescales by the EIC, SSC, and synchrotron processes at the external shock radius of $r={10}^{17.5}$~cm are shown. For comparison, the dynamical timescale is also shown. One can see that the corresponding $\gamma_{e,c} \sim {10}^{4.5}$. Source parameters are the same as those used in the caption of Figure~4.}
\end{figure}

The resulting spectra for our typical parameter sets are shown in Figures~3 and 4. 
As expected in the previous section, the EIC peak is located at $E_{\rm EIC}^c \sim 10-100$~GeV.
In our cases, the EIC peak energy is comparable to the SSC peak energy at $T \sim T_a$, which can be understood from $E_{\rm EIC}^c/E_{\rm SSC}^c \sim E^b/2 E^c$. When the EIC emission is dominant, its spectrum is roughly expressed by Equations~(21). (When the SSC emission is dominant, its spectrum is roughly expressed by Equation~(29).) As expected before, the KN suppression becomes important above $\sim 1-10$~TeV but it is difficult to be observed due to the EBL attenuation. 

 For these parameter sets of $\mathcal{E}_{\rm LP,X}^{\rm iso}/\epsilon_{ef} \mathcal{E}_k \sim 1$ and $p \sim 2.4$, we see that the EIC flux becomes larger than the SSC flux at $T \sim T_a$. 
This can be understood by comparing electron cooling timescales. An example of three timescales is shown in Figure~5, where we can see that $\gamma_{e,c} \sim {10}^{4-5}$ at $r \sim {10}^{17.5}~{\rm cm}$ (or $T \sim {10}^{3}$~s). When the EIC and SSC cooling times can be estimated in the Thomson limit, we obtain  
\begin{eqnarray}
\frac{t_{\rm EIC}^{-1}}{t_{\rm SSC}^{-1}} &\sim& 40~L_{\rm LP,48}|_{T_a^{\prime}} {(7g_p/2)}^{1-p} f_{ef}^{2-p} \epsilon_{ef,-2}^{1-p} \epsilon_{Bf,-3}^{2-p} \mathcal{E}_{k,53}^{-p/2} n_0^{(2-p)/2} \nonumber \\ 
&\times& {(1+Y)}^{3-p} {(1+z)}^{-p/2} T_3^{-\alpha_{\rm LP}+p/2}. \label{EICvsSSC}
\end{eqnarray}
Then, the ratio of the EIC flux to the SSC flux is roughly estimated as $\sim x t_{\rm SSC}/t_{\rm EIC}$. Note that, for sufficiently large $\gamma_{e,c}$ and/or $E^b$/$E^c$, the results are affected by the KN effect. 

In the case shown in Figure~5, the synchrotron cooling is dominant. If the EIC cooling is more important than the SSC cooling and the synchrotron cooling, the afterglow emission from the early jet is affected by the EIC cooling. However, this occurs only when the late prompt emission from the late jet is bright enough. The associated afterglow emission from the early jet, produced by electrons with $\sim \gamma_{e,c}$, is typically masked by the emission from the late jet, so that it seems difficult to observe the EIC influence at the optical or X-ray band.    

\begin{figure}
\includegraphics[width=\linewidth]{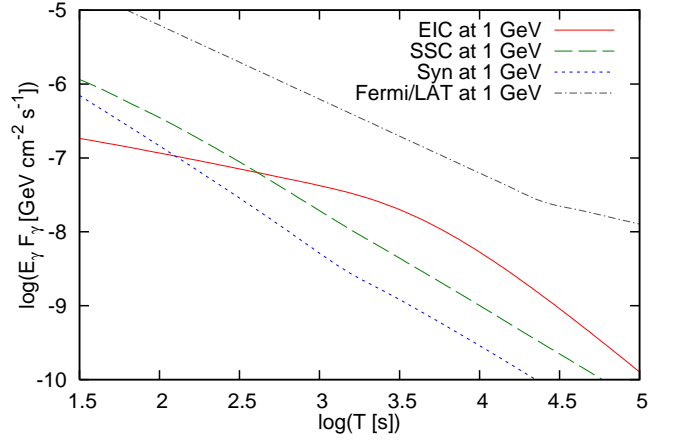}
\caption{Gamma-ray light curves of EIC emission at $1$~GeV in the late internal dissipation model for GRB afterglows, caused by Compton scatterings of X-ray photons by electrons accelerated at the external shock. For comparison, light curves of synchrotron and SSC emission are also shown.
The parameter set is the same as that used in the caption of Figure~3.
The {\em Fermi}/LAT sensitivity is overlaid. 
Note that the attenuation by pair creation both inside and outside the source is taken into account.}
\end{figure}
\begin{figure}
\includegraphics[width=\linewidth]{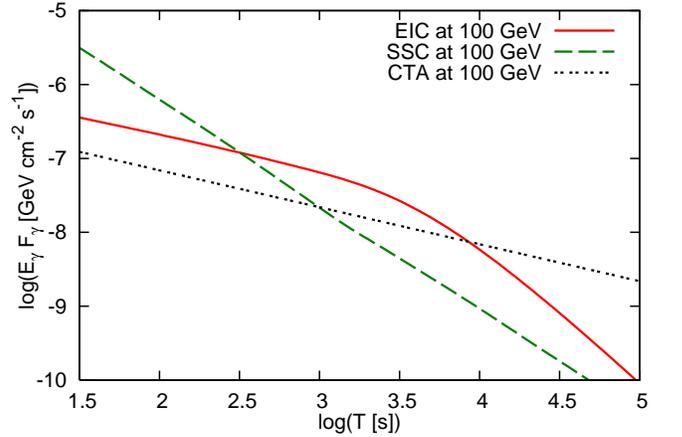}
\caption{Same as Figure~6, but at $100$~GeV. The CTA sensitivity (with the duty factor of 30~\%) is overlaid instead of the \textit{Fermi} one (CTA Consortium 2010).}
\end{figure}
\begin{figure}
\includegraphics[width=\linewidth]{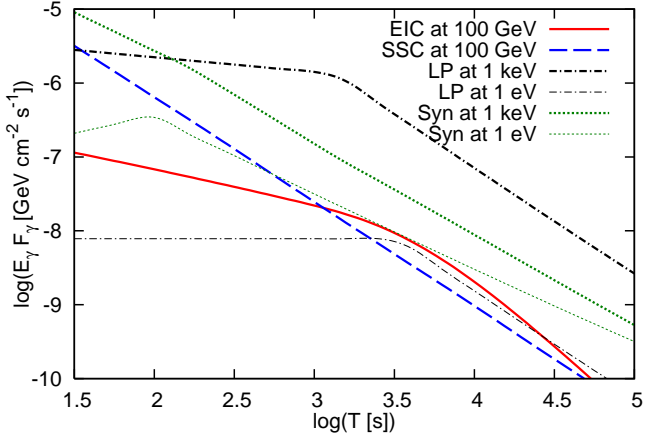}
\caption{Light curves of early and late jets in the late dissipation model at various energy bands. Syn and SSC come from synchrotron and SSC emission by relativistic electrons accelerated at the external shock of the early jet. LP represents the assumed seed photon emission from the late jet, which is responsible for shallow-decay X-ray emission, and EIC is the EIC emission by Compton scatterings of late prompt photons by electrons accelerated at the external shock. 
The parameter set is the same as that used in the caption of Figure~4.
Note that the attenuation by pair creation both inside and outside the source is taken into account.}
\end{figure}

The resulting light curves are shown in Figures~6, 7 and 8. 
The SSC flux evolves as $E F_{\rm SSC} \propto T^{-p/2}$ at $E > E_{\rm SSC}^c$. On the other hand, the EIC flux has shallower light curves, but its time evolution is different from that in the X-ray band (see Figure~8).  
In this sense, the EIC emission in the late internal dissipation model can be distinguished from the
predictions of other models, such as SSC emission from the late jet or SSC afterglow emission in modified FS models.  The time evolution of the EIC emission is understood from $Y_{\rm EIC}= 
t_{\rm EIC}^{-1}(\gamma_{e,c})/t_{\rm syn}^{-1}(\gamma_{e,c})$. If electrons with $\gamma_{e,c}$ are in the Thomson regime (which is not always true), we expect $Y_{\rm EIC} \propto L_{\rm LP}/R^2 \Gamma^2 B^2 \propto T^{-\alpha_{\rm LP}+1}$. On the other hand, the synchrotron luminosity in the slow cooling case obeys $L_{\rm SSC}^c \propto T^{-p/2}$ from Equation~(30). Then, we roughly expect $E_{\rm EIC}^c F_{\rm EIC}^c \sim x Y_{\rm EIC} (E^c F^c) \propto T^{-\alpha_{\rm LP}+1-p/2}$, which declines more rapidly than the shallow decay emission. For example, for $p \sim 2.4$ and $\alpha_{\rm LP} \sim 0.2$, we have $E F_{\rm EIC} \propto T^{-0.4}$. When the KN effect plays a role, the temporal index is somewhat steeper, which seems consistent with the numerical results.  
The break time of the shallow-decay emission is $T_a \sim {10}^{3}$~s, but the EIC flux does not decline for a while even after $T_a$. This is because seed photons interacting with FS electrons come from backward ($\theta \sim 0$), leading to suppressed EIC emission toward the observer, while photons passing through the FS with angles of $\theta \neq 0$, which significantly contribute to the EIC flux, are delayed compared to non-scattered photons from $\theta \sim 0$. 
Note that the timescale of this delay is of order of $R/\Gamma^2 c \sim T$, which is understood from the fact that EIC emission induced by an impulsive seed photon emission lasts until we observe photons entering the FS region with $\theta \sim 1/\Gamma$. 

From Figures~3 and 7, for our typical parameter sets, the EIC emission is expected at energies larger than 10~GeV. 
As we can see, the EBL attenuation is moderate for nearby GRBs, though it becomes crucial for distant bursts (see below).
At such very high energies, observations by Cherenkov telescopes such as MAGIC, VERITAS, HAWC, and CTA are more promising. Although no clear detections have been obtained so far~\cite{Abd+07,Alb+07,Aha+09,Ale+10}, future observations with HAWC and CTA would improve the chances for this, and either detections or non-detections are important to test the model.  
Detections by \textit{Fermi} are limited at late times, but they are being made in the earlier afterglow phase. However, note that the synchrotron or SSC emission is more important than the EIC emission at the earlier phase (especially just after the prompt emission), which can be expected from Equation~(40). Also, the SSC emission from the late jet, which has been discussed in the previous section, can be relevant in the GeV range (e.g., in the magnetic dissipation scenario), while the EIC emission and/or SSC emission from the early jet will be dominant in the 100~GeV range. 

\begin{figure}
\includegraphics[width=\linewidth]{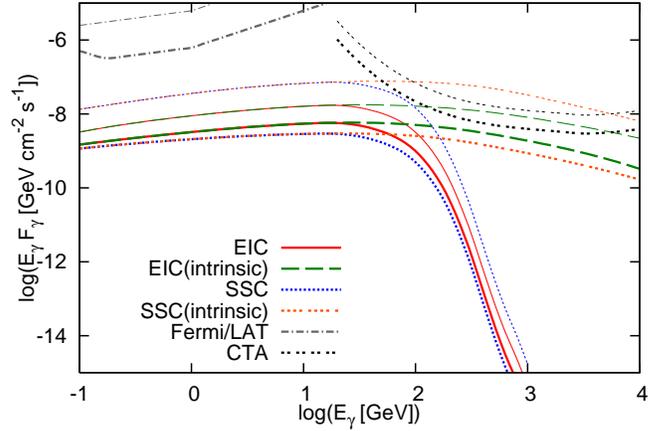}
\caption{Gamma-ray spectra of EIC and SSC emission in the late internal dissipation emission model at $T={10}^{2.6}$~s (thin) and $T={10}^{3.6}$ s (thick), but the source redshift is taken as $z=1$. Source parameters are the same as that used in the caption of Figure~3. 
The EBL attenuation is included in EIC and SSC, but not included in EIC (intrinsic) and SSC (intrinsic). One can see that it is crucial for detections by Cherenkov telescopes.}
\end{figure}
\begin{figure}
\includegraphics[width=\linewidth]{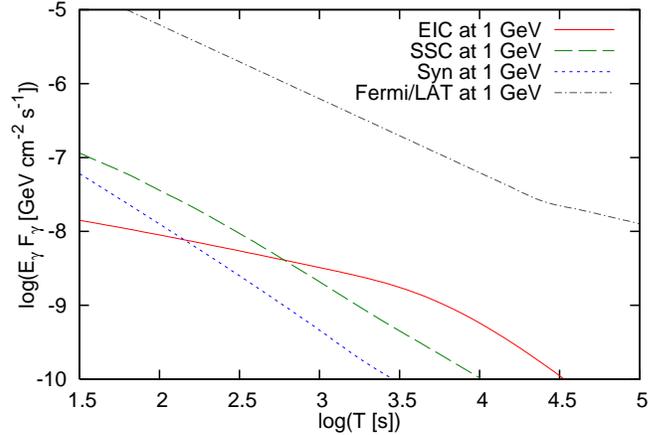}
\caption{
Gamma-ray light curves of EIC emission at $1$~GeV in the late internal dissipation model for GRB afterglows. For comparison, light curves of synchrotron and SSC emission are also shown.
The parameter set is the same as that used for Figure~9.
Here, the attenuation by pair creation both inside and outside the source is taken into account.}
\end{figure}
\begin{figure}
\includegraphics[width=\linewidth]{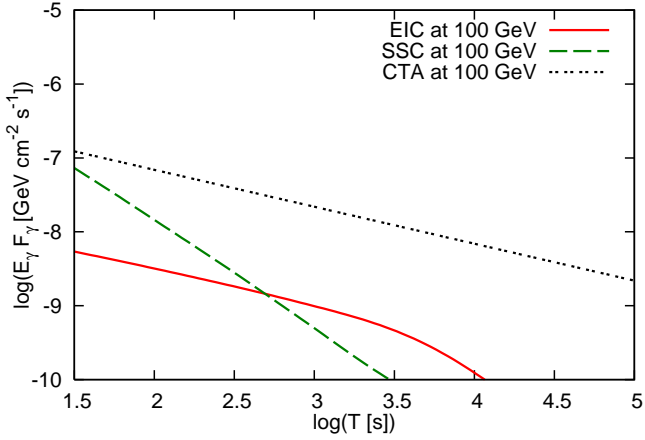}
\caption{
Same as Figure~10 but at $100$~GeV.}
\end{figure}

We have considered in our calculations the EBL attenuation, using the low-IR model developed by Kneiske et al. (2004).
Detecting gamma rays at very high energies above $100$~GeV is prevented by this EBL attenuation. Even for a burst at $z=0.3$, we have seen that the EBL attenuation largely degrades the resulting fluxes at $\gtrsim 300$~GeV. 
For a burst at higher redshifts, the situation becomes worse. In Figures~9, 10 and 11, results for $z=1$ are shown, where the EBL attenuation becomes crucial at $\gtrsim 100$~GeV. The EIC and SSC peak energies are higher than 
the cutoff by the EBL attenuation, so that both of the EIC and SSC fluxes are largely degraded. However, detections around $\sim 10$~GeV appear still promising at earlier times, even though they are difficult at late times.   

Note that gamma rays absorbed by the EBL must produce energetic pairs, which lead to IC emission by scatterings with the EBL photons. In our calculations, this pair echo emission is not included since it is beyond the scope of this work, although it could affect the observed afterglow emission if the intergalactic magnetic field in voids is weak enough \cite[e.g.,][]{RMZ04,Mur+09}. 

\subsection{Discussion on Parameter-Dependence}
We have demonstrated that the EIC emission dominates over the SSC emission in the late internal dissipation model for shallow-decay emission. 
Importantly, predictions of the EIC emission are straightforward, once X-ray and optical afterglows are well observed. The parameters necessary for calculations of high-energy emission are determined via fitting with the two-component (early and late) jet model as done in Ghisellini et al. (2009).
The parameter dependence of the relative importance of the EIC emission to the SSC emission is seen from Equation~(\ref{EICvsSSC}). The most important quantity is $\mathcal{E}_{\rm LP}^{\rm iso}/\epsilon_{ef} \mathcal{E}_{k}$ 
(which is expected by setting $p \sim 2$), which is seen by comparison between Figures~7 and 8. For our typical cases, the EIC emission is dominant at late times ($T \gtrsim T_a$), but can be less important for smaller values. In fact, there is large diversity among observed X-ray and optical afterglows so that it would be natural to expect that high-energy afterglows also exhibit a high diversity, depending on $\mathcal{E}_{\rm LP}^{\rm iso}/\epsilon_{ef} \mathcal{E}_{k}$. 

In our calculations, we have assumed $\epsilon_{Bf}={10}^{-3}$, but the EIC and SSC peaks are rather sensitive to 
$\epsilon_{Bf}$ (see Equations~(17) and (28)). We see that $E_{\rm EIC}^c/E_{\rm SSC}^c \propto \epsilon_{Bf}^{3/2}$, so 
that the EIC peak is more likely to be higher than the SSC peak for larger $\epsilon_{Bf}$. This implies that the EIC component is more frequently dominant over the SSC one at high energies. Note that the EIC peak energy can be around $1-10$~GeV rather than $0.1-1$~TeV when $\epsilon_{Bf} \sim {10}^{-2}$.  

Another potentially relevant parameter is $E^b$. For typical values used in this work, the results on the EIC emission are not so sensitive to this quantity, up to a modest factor (see Figure~12). But this may not be the case if the EIC cooling occurs in the KN regime. If $E^b$ is so large that the EIC cooling occurs in the KN regime while the SSC cooling does in the Thomson regime, the EIC emission would be more suppressed. 
So far, we have assumed that $E^b$ does not depend on time. Although this may not be true, it is difficult to determine its time evolution from observations. To see the influence of this uncertainty on results, we also calculate the EIC emission with the break energy of $E^b (T) \propto L_{\rm LP}^{1/2} (T)$. However, as seen from Figures~12 and 13, the results are hardly changed, because the EIC emission mainly occurs in the Thomson regime. At later times, the EIC flux with $E^b (T) \propto L_{\rm LP}^{1/2} (T)$ is a bit larger than that with $E^b=$const. This is because lower $E^b$ at late times can compensate the KN effect due to increase of $\gamma_{e,c}$.   

The low-energy photon index of the late prompt emission, $\beta_l$, is observationally uncertain. But this becomes crucial for the EIC spectrum at relatively low energies of $E \lesssim E_{\rm EIC}^m$, so that we expect that our results are not affected by this. In addition, the EIC flux is larger if $\beta_l > 1$.   

\begin{figure}
\includegraphics[width=\linewidth]{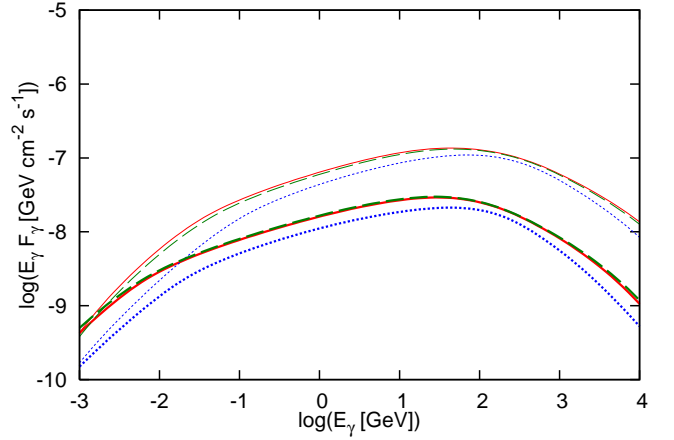}
\caption{EIC spectra calculated with different values of $E^b$. Intrinsic EIC spectra (where the EBL attenuation is not taken into account) are shown at $T={10}^{2.6}$~s (thin) and  $T={10}^{3.6}$~s (thick). The source redshift is set to $z=0.3$. 
For the solid curves, the used parameter set is the same as that in the caption of Figure~3. The dotted curves are for ${E'}^b={10}^{2.5}$~eV and the dashed curves are for ${E'}^b = 0.1~{\rm keV}~{(L_{\rm LP}^b/L_{\rm LP}^b|_{T_a^{\prime}})}^{1/2}$, while the other parameters are the same.}
\end{figure}
\begin{figure}
\includegraphics[width=\linewidth]{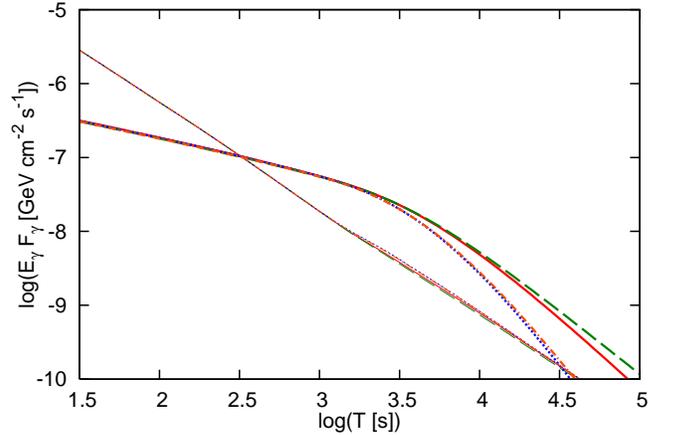}
\caption{EIC light curves (thick) at $100$~GeV, calculated with different assumptions on $E^b$ and $\Gamma_{\rm LP}$. For comparison, SSC light curves (thin) are also shown. The source redshift is set to $z=0.3$. 
For the solid curves, the used parameter set is the same as that in the caption of Figure~3. 
The dashed curves are for ${E'}^b = 0.1~{\rm keV}~{(L_{\rm LP}^b/L_{\rm LP}^b|_{T_a^{\prime}})}^{1/2}$ rather than ${E'}^b=$const. The dotted curves are for $\alpha_{\rm st}=2.8$ rather than $\alpha_{\rm st}=1.5$. The dot-dashed curves are for the case with the evolving ${E'}^b$ and $\alpha_{\rm st}=2.8$.
The attenuation by pair creation both inside and outside the source is taken into account.}
\end{figure}

We also show, for comparison, the resulting EIC and SSC light curves for $\alpha_{\rm st}=2.8$ in Figure~13, where one can see that the EIC light curve declines more rapidly after $T_a$.  For the jet opening angle and bulk Lorentz factor of the late jet, we have assumed $\Gamma_{\rm LP} > 1/\theta_{\rm LP}$. This may not be the case, as discussed in Ghisellini et al. (2007). If the late jet is decelerated with time, we expect that the observed light curve has the break when $\Gamma_{\rm LP}$ becomes $\sim 1/\theta_{\rm LP}$. This break may be the origin of $t_a$, although it is not clear why the late jet is decelerated continuously.  In this case, only a fraction of seed photons can interact with FS electrons after $T_a$. But this just corresponds to a change of $\alpha$, which is already taken into account in observable parameters.   

The jet opening angles and axes of the two jets are also assumed to be the same. However, we can still expect the EIC emission even if the jets are a bit misaligned. If either edge of the early jet is on the line of sight, photons from the late jet still come to the observer through the early jet (independently of the prompt emission mechanism), but the resulting EIC flux is reduced by a factor of two at most. 
Note that the important assumption used in this work is $(r_i/r) \Gamma \theta_{\rm LP} \ll 1$, which is typically valid in our model. When this condition does not hold, more detailed calculations are required.

\subsection{Specific Cases}
Bursts with a sudden decline in their X-ray afterglows are of particular interest.
For example, GRB 070110 has a steep decline of $\alpha_{\rm st}=9$ after the plateau of 
$\alpha_{\rm fl}=0.09$~\cite{Tro+07}. In Figures~14 and 15, we show the specific case of GRB afterglows with such a plateau (with $\alpha_{\rm fl}=0$ and $\alpha_{\rm st}=10$), to see the EIC emission induced by the plateau X-ray emission.  It is obvious that the EIC spectrum is similar to that shown in Figure~3, since a similar seed photon spectrum is assumed. 
On the other hand, the EIC light curve is different from those shown in Figure~4, reflecting different X-ray light curves. Before $T \sim T_a$, the EIC light curve is steeper than the X-ray one, as discussed before. However, this is not the case after $\sim T_a$. This is because the EIC emission is similar to high-latitude emission, so that the EIC emission does not show a sudden decline even though the seed photon emission ends abruptly. This was the behavior seen for an impulsive seed photon emission, as demonstrated in the prior emission model~\cite{Mur+10}. 
Detecting such a signature of high-latitude emission associated with the sudden decline after the 
plateau would be useful as evidence of the late internal dissipation model.   

\begin{figure}
\includegraphics[width=\linewidth]{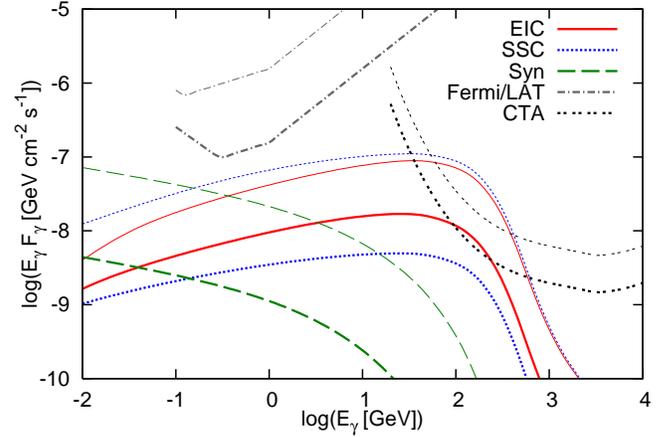}
\caption{Gamma-ray spectra of EIC emission from the GRB afterglow with the plateau X-ray emission at $T={10}^{2.6}$~s (thin) and $T={10}^{3.6}$~s (thick). For comparison, synchrotron and SSC emission from the standard afterglow component are also shown. The source redshift is set to $z=0.3$. 
Relevant parameters for the late jet are $L_{\rm LP}^b|_{T_a^{\prime}} = {10}^{48.5}~{\rm erg}~{\rm s}^{-1}$, $T_a^{\prime}={10}^{3}$~s, ${E'}^b=0.1$~keV, $\alpha_{\rm fl}=0$, and $\alpha_{\rm st}=10$. Relevant parameters for the standard afterglow component are the same as those in the caption of Figure~3. 
The attenuation by pair creation both inside and outside the source is taken into account.}
\end{figure}
\begin{figure}
\includegraphics[width=\linewidth]{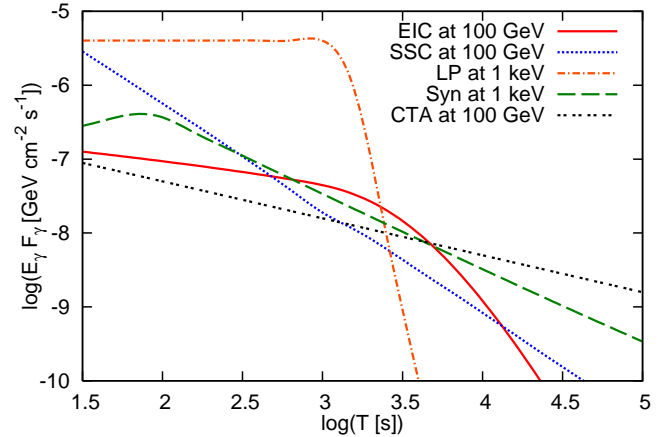}
\caption{Light curves of early and late jets in the late dissipation model at various energy bands. Syn and SSC come from synchrotron and SSC emission by relativistic electrons accelerated at the external shock of the early jet. LP represents the assumed seed plateau emission from the late jet, and EIC is the EIC emission by Compton scatterings of late prompt photons by electrons accelerated at the external shock. 
The parameter set is the same as that used for Figure~14. We can see that the EIC light curve is much shallower than that of late prompt emission after $T_a$. 
The attenuation by pair creation both inside and outside the source is taken into account.}
\end{figure}

Another case of interest is that of the GRBs that were observed by \textit{Fermi} which may be
represented by the late dissipation models. The high-energy emission detected by \textit{Fermi} may originate from the external shock. For example, Kumar \& Barniol Duran (2010) argued that long-lasting GeV emission comes from electrons accelerated at the FS caused by an adiabatic relativistic blast wave expanding into a low density ISM ($n \sim {10}^{-4}~{\rm cm}^{-3}$), with a low magnetic field ($\epsilon_{Bf} \sim {10}^{-4}$).  On the other hand, Ghisellini et al. (2010) argued that long-lasting GeV emission may be explained by a radiative relativistic blast wave, with $\epsilon_{ef} \sim 1$ and $p \sim 2$.  
Although the origin of GeV emission especially at the very early stage is still under debate~\cite{He+10,LW10}, the late-time GeV emission is likely to be regarded as afterglows. 

Just for demonstrative purposes, we also calculate the EIC emission for a burst like GRB 090902B. 
Unfortunately, there have been no bursts that have canonical early afterglow light curves simultaneously observed by \textit{Fermi} and \textit{Swift}, so that we just show the result for parameters provided in Cenko et al. (2010) in Figure~16 (see also Pandey et al. 2010) (but the redshift is set to $z=1$ here). Those parameters are indicated from late-time observations but not exact ones for explaining long-lasting GeV emission, since implied $\gamma_{e,c}$ is smaller than those used in Kumar \& Barniol Duran (2010). In fact, Liu \& Wang (2010) showed that the GeV emission may rather be explained by an additional jet component (although the two-component jet model used there is different from that considered here).   
Here, we are not pursuing possibilities to explain the long-lasting GeV emission with the late internal dissipation model, so these parameters are suitable enough for the present purpose. As can be seen, for our parameters on late prompt emission, the EIC component is dominant over the SSC one at late times. The EIC component is especially important at high energies of $\sim 10-100$~GeV. At lower energies, the synchrotron component dominates over the others, although the curve shown is fairly optimistic (since $\gamma_{e,M} = \sqrt{\frac{6 \pi e}{\sigma_T B \eta}}$ is used assuming that the upstream magnetic field is the downstream one and $\eta \sim 1$). As is demonstrated in Figure~16, the EIC emission induced by late prompt emission can typically be important in very high energies only at relatively late times, so that our results on high-energy afterglow emission are compatible with the \textit{Fermi} observations. 
On the other hand, high-energy late prompt emission could potentially be relevant in the GeV range. One may expect that it shows the shallow-decay behavior when the shallow-decay emission comes from the late internal dissipation. However, the observational situation is currently unclear since simultaneous detections by \textit{Fermi} and \textit{Swift} are required. Possibly, for bursts detected by \textit{Fermi}/LAT, $T_a$ is large enough and it may become important only at late times, or  $T_a$ is small enough but it may be masked. Or, declining high-energy emission could happen, if the steep-decay emission comes from the late internal dissipation. Also, \textit{Fermi}/LAT bursts tend to be most energetic ones, and it has not been settled whether accelerated electrons are in the fast or slow cooling regime~\cite{Ghi+10,KD10}. Future simultaneous detections of high-energy gamma rays from GRBs with canonical afterglow light curves are anticipated.

\begin{figure}
\includegraphics[width=\linewidth]{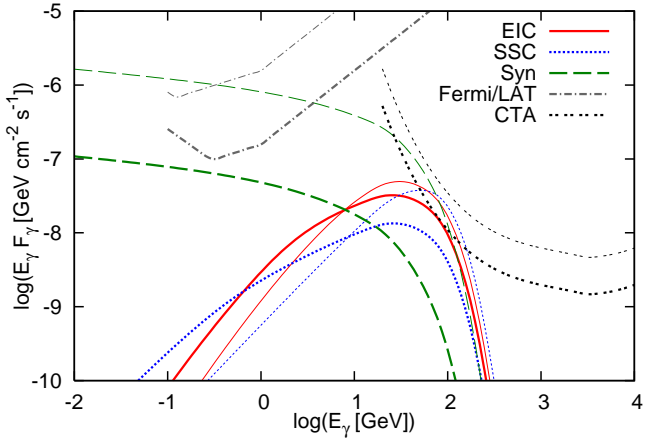}
\caption{Gamma-ray spectra of EIC emission for a burst with GRB 090902B-like afterglow parameters, at $T={10}^{2.6}$~s (thin) and $T={10}^{3.6}$~s (thick). For comparison, synchrotron and SSC emission from the standard afterglow component are also shown. The source redshift is set to $z=1$. 
Relevant parameters for the late jet are $L_{\rm LP}^b|_{T_a^{\prime}} =1.02 \times {10}^{50}~{\rm erg}~{\rm s}^{-1}$, $T_a^{\prime}={10}^{3}$~s, ${E'}^b=0.1$~keV, $\alpha_{\rm fl}=0$, and $\alpha_{\rm st}=2$. 
Relevant parameters for the standard afterglow component are $\mathcal{E}_{k}=6.8 \times {10}^{53}~{\rm erg}$, $\epsilon_{ef}=0.15$, $\epsilon_{Bf}=0.058$, $n=5.8 \times {10}^{-4}~{\rm cm}^{-3}$, and $p=2.22$. 
The attenuation by pair creation both inside and outside the source is taken into account.}
\end{figure}

\section{Summary and Discussion}
In this paper, we have studied the possibility that the shallow-decay or plateau emission 
originates from late internal dissipation in the late jet driven by the long-lasting central engine (e.g., mass fall back onto a black hole or rotational energy loss of fast rotating magnetars).  
We have discussed various theoretical scenarios of the emission mechanism, late internal shock, magnetic dissipation, and photospheric scenarios. There are few clues to the origin of the late prompt emission, and all the three scenarios seem compatible with observations at present.  

We have also investigated the associated high-energy emission in the late internal dissipation model and discussed two possibilities: high-energy late prompt emission and high-energy afterglow emission. The former comes from internal dissipation in the late jet, and the predictions depend on the specific scenarios. For example, in the photospheric scenario, high-energy emission may be produced by the IC process but $>$~GeV emission is not expected in the one-zone case due to the large pair-creation opacity around the photosphere. On the other hand, the late internal shock and magnetic dissipation scenarios may lead to $\sim 1-10$~GeV gamma rays by the SSC mechanism, which are important for \textit{Fermi}. As demonstrated in this work, detections by \textit{Fermi} and possibly CTA are expected for nearby and/or energetic GRBs, which would be useful for revealing the mechanism of late prompt emission.  

The latter possibility includes the SSC and EIC emissions produced by electrons accelerated at the external shock, which will be especially relevant in the very high energy range. Especially, the EIC emission, which is high-energy afterglow emission induced by late prompt photons, is not so sensitive to details of models and should be useful as a test of the existence of late internal dissipation during the shallow-decay phase.  
In this work, we have investigated the EIC emission both analytically and numerically and demonstrated that the EIC flux may become larger than the SSC flux around the end time of the shallow-decay phase. The EIC peak is typically expected at $\sim1-100$~GeV, and the EIC emission typically has a steeper light curve than in the X-ray one, but a shallower one when the X-ray light curve shows a sudden decline. Hence, it would be possible to distinguish it from the other possibilities such as SSC components from the early and late jets. 
We also expect that it is easier for the synchrotron and SSC components to dominate at very earlier times.  

Although the detectability depends on the parameters and on the EBL, ground-based gamma-ray 
observatories, such as MAGIC, VERITAS, HESS, CTA and HAWC, would be important tools in the search for such signals. Very high energy gamma rays from GRBs have not been firmly observed so far~\cite{Abd+07,Alb+07,Aha+09,Ale+10} and the event rate of nearby bursts is not large. (For example, the rate of GRBs  occurring at within $z \sim 0.3$ is estimated as $\sim$~a few events per year~\citep[e.g.,][]{LZVD07}.) Nevertheless, once sufficiently fast follow-up observations are successful for nearby events, Cherenkov telescopes with a low-energy threshold ($\sim 30$~GeV) may allow us to have good photon statistics thanks to their high sensitivities. 
Theoretical predictions of the EIC emission are testable once parameters are specified from observations, and the strategy for testing the model is as follows.
First, one determines the relevant standard afterglow parameters for the early flow. 
When afterglows are well observed at optical (and/or X-ray) bands, the parameters such as $\mathcal{E}_k$, $p$, $\epsilon_{Bf}$ and $\epsilon_{ef}$ are determined in the context of the standard external FS theory.
At the same time, parameters on the late prompt emission, such as $L_{\rm LP}^b$ and $E^b$ and $T_a$, can also be determined from observations at X-ray (and/or optical) bands.  With those parameters, both the EIC and SSC emissions are calculated and can be compared to high-energy observations. Even non-detections would provide useful constraints on the models, especially for GRBs with a strong plateau or shallow-decay emission. 
 
It is also important to keep in mind that GRB afterglows seem to be fairly diverse~
\citep[e.g.,][]{Ghi+09,Lia+09}. Although GRBs with a shallow-decay emission may be explained by a late jet from the long-lasting central engine, some GRBs do not show the shallow decay and can be explained by the standard afterglow model, where the synchrotron or SSC emission is expected to be dominant. 
For this reason, multi-wavelength observations from radio to gamma rays are important for comprehensive studies of GRB afterglows. 

\section{acknowledgments} 
K.M. acknowledges financial support by a Grant-in-Aid from JSPS, from CCAPP and from PSU. 
K.T. and P.M. acknowledge partial support from NASA NNX08AL40G, NASA NNXAT72G, 
NSF PHY-0757155 and U.R.A. 10-S-017.
This research was also supported by Grant-in-Aid from the Ministry of Education, Culture, Sports, Science and Technology MEXT of Japan, no.~19047004 and 21740184 (R.Y.). 
The numerical calculations were carried out on Altix3700 BX2 at YITP in Kyoto University.
P.M. acknowledges the hospitality of the Institute for Advanced Study, Princeton, during
part of this project.

\clearpage



\appendix
\section{Formulas of EIC Emission}
Here, we derive formulas of EIC emission in GRB afterglows. 
The observed flux from the shell expanding toward us relativistically is~\citep[e.g.,][]{GPS99,WL99}  
\begin{equation}
F (T) = \frac{1+z}{d_L^2} \int d \phi \, \int d \cos \theta \, \int d r \, r^2 \frac{\tilde{j}_{\varepsilon}}{\Gamma^2 {(1-\beta \cos \theta)}^2},
\end{equation}
where $\tilde{j}_{\varepsilon}$ is the comoving emissivity and $\varepsilon$ is the seed photon energy in the comoving frame. Hereafter, we also use $\tilde{E}=(1+z)E \Gamma (1-\beta \cos \theta)$ and $T=(1+z)(\hat{T}-r \cos \theta/c)$. Especially, the comoving EIC emissivity is written as~\citep[e.g.,][]{TWM09}
\begin{equation}
\tilde{j}_{\varepsilon} = \frac{3}{2} \sigma_T (1- \cos \tilde{\theta}) \int d \gamma_e \, \frac{d n_e}{d \gamma_e} \int dy \, \tilde{J}_{\varepsilon}^{\rm seed} (1-\xi) 
\left[ 1-2 y + 2 y^2 + \frac{\xi^2}{2(1-\xi)}\right],
\end{equation}
where
\begin{equation}
\tilde{J}_\varepsilon^{\rm seed} = \frac{1}{2\Gamma} \left(\frac{1}{4 \pi r^2} \frac{d_L^2}{1+z} F_{\rm seed} \right).
\end{equation}
Here, $y \equiv \frac{\xi m_e c^2}{2(1-\cos \tilde{\theta}) \gamma_e \varepsilon (1-\xi)}$, $\xi \equiv \frac{\tilde{E}}{\gamma_e m_e c^2}$, and scattering angles $\theta$ and $\tilde{\theta}$ are measured in the central engine frame and the comoving frame, respectively. The range of $y$ is $\frac{1}{2 (1-\cos \tilde{\theta}) \gamma_e^2 (1-\xi)} \leq y \leq 1$. Note that Equation~(A2) is easily obtained from
\begin{equation}
\frac{d N_{\rm EIC}^{(1)}}{d \tilde{E} d \tilde{T} d\tilde{\Omega}} \approx \frac{3}{16 \pi \gamma_e^2} \sigma_T  c \int d \varepsilon \, \frac{1}{\varepsilon} \frac{d n_{\rm seed}}{d \varepsilon}\left[ 1-\frac{2 \xi}{b_{\tilde{\theta}} (1-\xi)}  +\frac{2 \xi^2}{b_{\tilde{\theta}}^2 {(1-\xi)}^2} +\frac{\xi^2}{2(1-\xi)} \right]
\end{equation}
where $b_{\tilde{\theta}} = 2(1-\cos \tilde{\theta}) \gamma_e \varepsilon/m_e c^2$~\cite{AA81,FP08}. 

First, we shall derive the formula for an impulsive seed photon spectrum. We also assume that seed photons come from $r_i \ll r$. In the case of instantaneous emission (at $t_0$) from an infinitely thin shell (at $R_0$), by using the replacement of $\tilde{j}_{\varepsilon} \rightarrow \tilde{j}_{\varepsilon}~\delta(\hat{T}-\hat{T}_0)~t_{\rm dyn}~\delta (r-R_0)~\tilde{\Delta}$, we obtain
\begin{eqnarray}
F_{\rm EIC} (T) &=& \frac{3}{2} \sigma_T (1-\cos \tilde{\theta}) \int 
d \gamma_e \, \frac{d n_e}{d \gamma_e} \frac{\tilde{\Delta}}{\kappa} 
\int d y \, \frac{\bar{F}_{\rm seed} |_{T_0}}{{(1+\Gamma^2 \theta^2)}^2}
(1-\xi) \left[ 1- 2 y +2 y^2 + \frac{\xi^2}{2(1-\xi)} \right]
\end{eqnarray}
where $t_{\rm dyn}=\tilde{\Delta}/c$ is the comoving dynamical timescale and $\tilde{\Delta} = R_0/\kappa \Gamma$ is the comoving shell thickness. Here, 
\begin{eqnarray}
\theta^2 (T) = 2 \left[ 1-\frac{c}{R_0} \left( \hat{T}_0-\frac{T}{1+z} \right) \right]
\end{eqnarray}
In the case of a broken power-law seed spectrum, we can write $\bar{F}_{\rm seed}|_{T_0} \equiv \bar{F}_{\rm seed}^b |_{t_0} G (\varepsilon)$, where $\bar{F}_{\rm seed}^b |_{T_0} = \frac{L_{\rm seed}^b 2 \Gamma \Delta T}{4 \pi d_L^2 E^b t_{\rm dyn} (1+z)}$ which is smeared over the dynamical timescale of the shell. Note that $\Delta T$ is the duration of impulsive seed photon emission in the oberver frame. Equation~(A5) is essentially the same as Equation~(5) used in Murase et al. (2010)~\footnote{There was an unimportant typo in that paper, but calculations were performed using the correct expression, dropping off $\kappa$.}.

Next, we shall derive the formula for continuous seed photon emission. This is obtained by the similar procedure. Performing the replacement of $\tilde{j}_{\varepsilon} \rightarrow \tilde{j}_{\varepsilon}~\delta (\tilde{r}-\tilde{R}(\hat{T}))~\tilde{\Delta}(\hat{T})$ leads to
\begin{equation}
F_{\rm EIC} (T)=\frac{3}{2} \sigma_T \int d r \, (1-\cos \tilde{\theta}) \int d \gamma_e \, \frac{d n_e}{d \gamma_e} \tilde{\Delta} \int d y \, (1-\xi) \left[ 1- 2 y +2 y^2 + \frac{\xi^2}{2(1-\xi)} \right]
\frac{1}{4 r \beta} \frac{F_{\rm seed}^b (T) G (\varepsilon)}{\Gamma^4 {(1-\beta \cos \theta)}^2},
\end{equation}
where $\tilde{\Delta}=r/\kappa \Gamma$ and $\theta=\theta(r)$ is given by
\begin{equation}
\cos \theta = \frac{c}{r} \left(\int^r dr \, \frac{1}{c \beta} - \frac{T}{1+z} \right),
\end{equation}
and $\tilde{\theta}=\tilde{\theta}(r)$ is obtained via the Lorentz transformation. When $\Gamma \theta \gg 1$, we obtain Equation~(\ref{EICformula}).

\section{Distribution of Nonthermal Electrons}
In order to calculate both the EIC and SSC emission, we use the following electron distribution for $\gamma_{e} \geq \gamma_{e,m}$, which would approximately mimic the distribution of relativistic electrons in the dynamical timescale, 
\begin{equation}
\frac{d n_e}{d \gamma_e} \propto {\rm min} [1, f_{\rm cool}^{-1}] \gamma_e^{-p}, 
\end{equation}
where $p$ is the spectral index of accelerated electrons and $f_{\rm cool} \equiv t_{\rm dyn}/t_{\rm cool}$ is the effective optical depth for energy losses. In the slow cooling case with $t_{\rm cool}=t_{\rm syn}$ (where $t_{\rm syn}$ is the synchrotron cooling timescale), we have $d n_e/d \gamma_e \propto \gamma_e^{-p}$ for $\gamma_{e,m} \leq \gamma_{e}<\gamma_{e,c}$ and $d n_e/d \gamma_e \propto \gamma_e^{-p-1}$ for $\gamma_{e} \geq \gamma_{e,c}$. In the fast cooling case, we set $p=1$ for $\gamma_{e,c} \leq \gamma_e < \gamma_{e,m}$, which reproduces $\propto \gamma_e^{-2}$ if $t_{\rm cool}=t_{\rm syn}$. 
The value of $\gamma_{e,c}$ is determined by finding solutions of~\citep[e.g.,][]{NAS09,Wan+10} 
\begin{equation}
t_{\rm dyn}^{-1} = t_{\rm syn}^{-1}(\gamma_e)+t_{\rm SSC}^{-1}(\gamma_e)+t_{\rm EIC}^{-1}(\gamma_e), 
\end{equation}
where the IC loss timescales are evaluated from 
\begin{equation}
t_{\rm IC}^{-1} = \frac{c \gamma_e}{(\gamma_e-1)} \int d \mu \, (1-\mu) \int d\varepsilon \, \frac{d n_{\rm seed}}{d \varepsilon d \mu} (K_{\rm IC} \sigma_{\rm IC}) 
\end{equation}
where $K_{\rm IC}$ is the electron inelasticity for the IC process (which is calculated from Equation~(C2)) and $\sigma_{\rm IC}$ is the IC cross section which is given by the KN formula. 

The normalization is determined by 
\begin{equation}
\int d \gamma_e \, \frac{d n_e}{d \gamma_e} (4 \pi r^2 \tilde{\Delta}) = \mathcal{N}_e = \frac{4 \pi}{3-k} n r^3, 
\end{equation}
where $k=0$ for the ISM and $k=2$ for the wind medium. (A somewhat different normalization, $\mathcal{N}_e \approx 4 \pi r^2 (4 \Gamma n) (r/4 \Gamma)$ was used in Murase et al. 2010.) 

\section{SSC Emission and Pair Production}
In this work, we also calculate the SSC emission for comparison. For simplicity, we simply calculate the observed SSC flux from the comoving SSC power per comoving energy. 
The comoving SSC power per comoving energy is given by~\cite{BG70}  
\begin{equation}
\tilde{E} \frac{d N_{\rm SSC}}{d \tilde{E} d \tilde{T}} = \int d \gamma_e \frac{d \mathcal{N}_e}{d \gamma_e} \int d \varepsilon \frac{d n_{\rm syn}}{d \varepsilon} \tilde{E} \left< \frac{d {\sigma}_{\rm IC}}{d \tilde{E}} c' \right>
\end{equation}
where $c'=c(1-\mu)$ and
\begin{equation}
\left< \frac{d \sigma_{\rm IC}}{d \tilde{E}} c' \right> = \frac{3}{4} \sigma_T c \frac{1}{\gamma_e^2 \varepsilon} \left[ 1+v - 2v^2 + \frac{v^2 w^2 (1-v)}{2(1+vw)}+ 2v \ln v  \right],
\end{equation}
and $v \equiv \frac{\tilde{E}}{4 \varepsilon \gamma_e^2 (1-\xi)}$ and $w \equiv \frac{4 \varepsilon \gamma_e}{m_e c^2}$. Note that numerically calculated SSC fluxes have convex curves, which lead to larger fluxes compared to analytically calculated SSC segments~\cite{SE01}, and the Klein-Nishina effect becomes often important above the TeV range~\citep[e.g.,][]{Wan+10}. 
As for the seed photon spectrum, the analytical synchrotron spectrum is used in this work, which is expressed as three segments both in the fast and slow cooling cases~\cite{SPN98}. 

High-energy gamma rays may suffer from pair-production process with target photons in the source. We also take into account the resulting gamma-ray attenuation in the source. The optical depth for the pair production is expressed as
\begin{equation}
\tau_{\gamma \gamma} (\tilde{E}) = \frac{\tilde{\Delta}}{2} \int d \mu \, (1-\mu) \int d \varepsilon \, \frac{d n_{\rm syn}}{d \varepsilon} \frac{3}{16} \sigma_T (1-\beta_{\rm CM}^2) \left[2 \beta_{\rm CM} (\beta_{\rm CM}^2-2) + (3-\beta_{\rm CM}^4) \ln \left( \frac{1+\beta_{\rm CM}}{1-\beta_{\rm CM}} \right) \right], 
\end{equation}
where $\beta_{\rm CM}=\sqrt{(1-4 m_e^2 c^4/S)}$ and $S$ is the Mandelstam variable. In this work, pair attenuation in the source is taken into account by introducing the suppression factor $1/(1+\tau_{\gamma \gamma})$~\cite{Bar06}. More detailed discussions on the opacity effect is found in Granot et al. (2008). When the gamma-ray attenuation becomes crucial, one has to consider electromagnetic cascades in the source. However, as long as we consider the afterglow emission in the late phase as done in this work, the gamma-ray attenuation is not important, and we can neglect the electromagnetic cascades caused by the leptonic SSC and EIC emission. The situation is different when one consider hadronic scenarios, where hadronic cascades may be important at very high energies~\cite{PW05}.  



\end{document}